\documentclass[preprint,12pt]{elsarticle}

\usepackage{graphicx}
\usepackage{amsmath}
\usepackage{commath}
\usepackage{mathtools}
\usepackage{amssymb}
\usepackage{lineno,hyperref}
\usepackage{algorithm}
\usepackage{algpseudocode}

\usepackage{lineno}
\usepackage{subfigure}
\usepackage{bm}
\usepackage{cleveref}
\usepackage{graphicx}
\usepackage{booktabs}
\usepackage{siunitx}
\usepackage[T1]{fontenc}

\begin{document}

\begin{frontmatter}


\hypersetup{pdfauthor={Name}}
\title{Characterization and Generation of 3D Realistic Geological Particles with Metaball Descriptor based on X-Ray Computed Tomography}


\author[inst1,inst2,inst3]{Yifeng Zhao}

\author[inst2,inst3]{Xiangbo Gao}

\author[inst2,inst3]{Pei Zhang}

\author[inst3]{Liang Lei}



\author[inst2,inst3]{S.A. Galindo-Torres \corref{cor1}}
\cortext[cor1]{corresponding author}
\ead{s.torres@westlake.edu.cn}

\author[inst3]{Stan Z. Li \corref{cor2}}
\cortext[cor2]{Co-corresponding author }
\ead{Stan.ZQ.Li@westlake.edu.cn}

\affiliation[inst1]{organization={Collage of Environmental and Resources Science, Zhejiang University},
            addressline={866 Yuhangtang Road}, 
            city={Hangzhou},
            postcode={310058}, 
            state={Zhejiang Province},
            country={China}}

\affiliation[inst2]{organization={Key Laboratory of Coastal Environment and Resources of Zhejiang Province, School of Engineering, Westlake University,},
            addressline={18 Shilongshan Road}, 
            city={Hangzhou},
            postcode={310024}, 
            state={Zhejiang Province},
            country={China}}

\affiliation[inst3]{organization={School of Engineering, Westlake University,},
            addressline={18 Shilongshan Road}, 
            city={Hangzhou},
            postcode={310024}, 
            state={Zhejiang Province},
            country={China}}

\begin{abstract}
The morphology of geological particles is crucial in determining its granular characteristics and assembly responses. In this paper, Metaball-function based solutions are proposed for morphological characterization and generation of three-dimensional realistic particles according to the X-ray Computed Tomography(XRCT) images. For characterization, we develop a geometric-based Metaball-Imaging algorithm. This algorithm can capture the main contour of parental particles with a series of non-overlapping spheres and refine surface-texture details through gradient search. Four types of particles, hundreds of samples, are applied for evaluations. The result shows good matches on key morphological indicators(i.e., volume, surface area, sphericity, circularity, corey-shape factor, nominal diameter and surface-equivalent-sphere diameter), confirming its characterization precision. For generation, we propose the Metaball Variational Autoencoder. Assisted by deep neural networks, this method can generate 3D particles in Metaball form, while retaining coessential morphological features with parental particles. Additionally, this method allows for control over the generated shapes through an arithmetic pattern, enabling the generation of particles with specific shapes. Two sets of XRCT images different in sample number and geometric features are chosen as parental data. On each training set, one thousand particles are generated for validations. The generation fidelity is demonstrated through comparisons of morphologies and shape-feature distributions between generated and parental particles. Examples are also provided to demonstrate controllability on the generated shapes. With Metaball-based simulations frameworks previously proposed by the authors, these methods have the potential to provide valuable insights into the properties and behavior of actual geological particles.

\end{abstract}


\end{frontmatter}


\section{Introduction}
The geological particle (e.g. sands, cobblestones, rocks, etc.) has garnered significant attention in recent decades within the fields of geology and geotechnics, due to its significance in a wide range of physical and industrial processes\cite{meng2020three, golombek2020geology, zhang2021improved, tolomeo2022modelling}. It has been suggested that the macro-behavior of granular media (e.g. soils, minerals etc.) is closely connected to the micro-structural features of contained particles, where the morphology is one of the most dominant factors\cite{altuhafi2016effect, santamarina2004soil, zhao2021grain, lu2019re, chen2020effect}. As illustrated by various observations and theories, the particle shape governs not only granular characteristics\cite{shinohara2000effect, xiao2019effect, zhang2022size} containing friction, interaction and deformation but also assembly responses\cite{zuo2019experimental, yin2020effect, zhou2020study} enclosing permeability, strength and failure. 


On elucidating the impact of particle shape on the granular soil, simulation schemes based on micro-mechanical models, especially the discrete element method (DEM)\cite{cundall1979discrete}, have prevailed in the past few decades. In original DEM, the particle is simplified as circles or spheres, which is hard to reflect the impact of shape, e.g. the resistance to rolling. Under this context, the key issue is how to fully reconstruct the morphology of granular matters in DEM simulations. The initial attempt starts from the rolling resistance model\cite{iwashita1998rolling, wensrich2012rolling}, which improves the contact model with friction mechanism from the surface roughness to mimic the particle shape. In spite of progresses in dealing with physical phenomenon such as inertial friction\cite{wensrich2012rolling} and shear banding\cite{coetzee2016calibration}, it still suffers from the introduction of a number of free parameters and the property of unrealistic fabric\cite{jiang2005novel}. The clustering technique is another interesting approach, which represents shape characteristics by clustering simple, regular graphics like spheres\cite{ferellec2010modelling}, Polyhedrons\cite{hohner2012numerical}, Ellipsoids\cite{regueiro2014micromorphic} and Sphere-polyhedrons\cite{galindo2013coupled}. This technique is widely used due to its simplicity yet with the drawback of impractical surface roughness\cite{grabowski2021comparative} and can be computationally expensive\cite{garcia2009clustered}. 

With the inspiring development of X-ray Computed Tomography (XRCT) and computer vision techniques, opportunities are provided to bring more accurate and sophisticated shape features into DEM simulations. Various methods are developed to compress realistic particle morphologies into a uniform mathematical representation. A good example is the Fourier descriptor, which is developed to capture particle shapes based on the average normalized Fourier spectrum of main contours from the targeted particle\cite{thomas1995use, mollon2013generating}. Through similar pattern, the Spherical-Harmonic (SH) descriptor can also be used to characterize shapes features (including non-convex ones)\cite{zhou2018three, zhou2015micromorphology, su20183d}. However, the above two methods are limited to tackling star-like particles, of which all line segments between particle-center and particle-surface points are located within the particle body\cite{su20183d}. Many particles, such as lunar soils and concave sand, do not follow such constraints. To overcome the above limitation in representation, Vlahini{\'c} et al\cite{vlahinic2017computed} apply the level-set function as the descriptor. However, the shape descriptor has to cooperate with mechanical models in practical simulations. It is necessary to consider its coupling scheme when using a descriptor. Although level-set function has an excellent characterization fidelity, it still has a reliance on computational resources due to the look-up table mechanism in coupling mechanical models (mainly in contact detection of DEM)\cite{kawamoto2016level, medina2019geometry, zhao2019poly}. To relieve the efficiency problem in coupling, Zhao and Zhao\cite{zhao2019poly, zhao2020universality} developed a poly-superellipsoid based descriptor yet at the cost of certain constraints on the expressed shape(such as smoothness and complexity). Recently, the Metaball descriptor is introduced by the authors\cite{zhang2021metaball} to reconstruct non-spherical particle shapes. With proper function form, the contact detection of it can be tackled at a low cost, which enables a more efficient simulation framework\cite{zhang2022coupled}. Such framework is further coupled with Lattice Boltzmann Method for simulations of more complicated physical processes in fluid-particle systems\cite{zhao2022metaball}. However, it remains difficult for Metaballs to characterize complex-shaped granular particles. Therefore, there is still a pressing need for a shape characterization framework with descriptor of better balance between the imaging quality of particle shape and computational efficiency in coupling mechanical models, especially for complex-shaped particles of angular or concave features. 



Apart from characterization, particle generation is another non-negligible challenge. Although XRCT can be used to scan all involved particles in application, particle scanning and image processing can be economy-costly and time-consuming. Actually in practical engineering, only a small number of particles (less than 10\% as reported in \cite{nie2020probabilistic}) can be scanned. Direct simulation with them will suffer from repetitive particle morphologies. This makes it necessary to generate realistic particles with coessential morphological features. As the development of the aforementioned shape descriptors, many attempts have been inspired to tackle generation tasks. Among them, the SH-based technique is a popular choice\cite{liu2011spherical}. It first incorporates geometric features into specific SH coefficients. Then, different algorithms like random field \cite{grigoriu2006spherical}, fractal dimension\cite{wei2018generation}, Nataf transformation\cite{nie2020probabilistic}, and principal component analysis\cite{xiong2021gene, zhou2017generation} are applied on the distilled SH coefficient to add small variances following the morpholigical pattern of parental particles for generation. The Fourier-descriptor based approach is similar implementation\cite{mollon2013generating, mollon2012fourier, mollon20143d, chen2022modified}. However, the above schemes suffer from some problems, including:
(1) underfitting and overfitting problems on shape-feature distributions of the generated particles, e.g. the distributions of surface area and volume\cite{wei2018generation, xiong2021gene}, (2) intractable problems in obtaining particles with specific morphological features\cite{buarque2018granular}, 
(3) involving complex mixture models, which are costly in computations and require specialized personal\cite{shi2021randomly}
and (4) requiring bridging or transformation into other descriptors before practical simulations, which often results in a trade-off between accuracy and efficiency\cite{lai2022signed}. Another interesting attempts are geometry-based algorithms\cite{jerves2017geometry, medina2019geometry, buarque2018granular, macedo2023shape}. It requires some shape feature distributions as "morphological DNA". Then, geometric stochastic cloning algorithms\cite{jerves2017geometry, medina2019geometry} or the genetic algorithm\cite{buarque2018granular, macedo2023shape} are implemented to generate particles based on those distilled DNA. Although with advance in performance, this type of method still has an reliance on the computational and human resources, which can not be obtained easily by all individuals or institutions. A framework with a more flexible shape descriptor, which can tackle the above problems, is therefore needed. 

Deep learning techniques, which are able to extract and conclude general features from high dimensional data, provide an opportunity to tackle generation problems\cite{bourilkov2019machine}. Among them, the variational autoencoder(VAE) is a popular choice\cite{kingma2013auto}. It achieves generative modeling through establishing a mapping between some latent variables and the generative targets\cite{higgins2016beta}. With neural-network based variational inference, it can interpret and learn the underlying casual relations within the data. And practical generation can then be carried out by controlling those latent variables. Note that those latent variables are regularized, which means variables mapped with similar generated results are located together. This can brings many advantages, including feature extraction, generation quality and generation control\cite{tan2018variational}. Many successful applications have verified its superiority in generation tasks over traditional methods\cite{shi2021randomly, tan2018variational}.



On the above dilemmas in characterization and generation of geological particles, this paper presents Metaball-based solutions. For characterization, we propose a geometric-based Metaball-Imaging (MI) algorithm, which can capture complex particle morphologies directly from XRCT images with the Metaball descriptor. The utilization of geometric constraints allows Metaballs to capture more shape-details, even angular edges and concave voids. Four types of particles in distinct morphologies are selected for validation. The impact of the control point number, is carefully investigated. On this basis, the abilities of MI to reconstruct single particle and particle groups are examined at length on major morphological features, including the volume, area, sphericity, circularity, Corey shape factor, nominal diameter and volume-equivalent diameter. For generation, we develop the Metaball-based Variational Autoencoder (MetaballVAE). It can learn from the XRCT image of targeted grains and generate random Metaball-based particles retaining major morphological features from a regularized latent space. Note that MetaballVAE can learn not only the geometric features of the single particle but also feature distributions of the particle group. The regularized latent space makes it possible to modify particle morphologies in an arithmetic pattern, making it possible to obtain particles with specific shapes. Examined with two groups of particles different in sample number and geometric characteristics, the proposed generation algorithm has proved to be robust and effective.


\section{Characterizing 3D Realistic Granular Particles with XRCT images based on Metaball Descriptor}\label{MI}
\subsection{The Metaball descriptor}\label{metaball}
\begin{figure}
    \begin{centering}
    \includegraphics[width=1.0\linewidth]{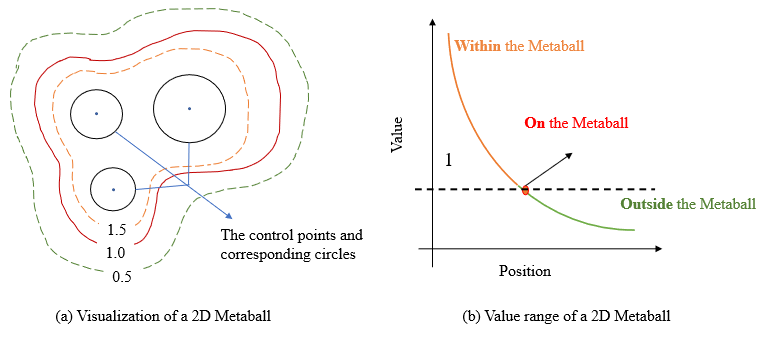}
    \caption{The visualization and value range of a 2D Metaball. The black points stand for the control points. And the black circles are the corresponding control circles. }
    \label{fig:Metaball2d}
    \end{centering}
\end{figure}

The Metaball function is a shape descriptor firstly proposed by Blinn\cite{blinn1982generalization}. It is expressed as a function of a distance between the probing point and a number of control points. Herein, we choose the typical inverse-square form:
\begin{equation}
    f(\boldsymbol{x})=\sum_{i=1}^{n} \frac{\hat{k_{i}}}{\left(\boldsymbol{x}-\boldsymbol{\hat{x}_{i}}\right)^{2}}
    \label{metaballM}
\end{equation}
where $n$ is the number of control points. $\boldsymbol{\hat{x}_{i}}$ is the position of control point $i$, representing a skeleton for the parameterized shape. $\hat{k_i}$ is the positive coefficient determining the weight of $i$th control point. Together, a pair of $\hat{k_i}$ and $\boldsymbol{\hat{x}_{i}}$ constitutes a control circle in 2D (black circles in Figure \ref{fig:Metaball2d}, a) or sphere in 3D. By setting this function equal to value 1, these control circles or spheres form the contour of the targeted particle surface, of which the parameter is noted as $\boldsymbol{M}$=$\{\hat{k_i}, \boldsymbol{\hat{x}_{i}}\}$. A 2D Metaball example is shown in Figure \ref{fig:Metaball2d} (a), where the particle surface contour is represented by red line.


Note that the value range of the Metaball function with random input is from $0$ to $\infty$. For 2D Metaball as shown in Figure \ref{fig:Metaball2d} (b), the function value will be exactly 1 at the surface, with lower values for points outside the particle and higher values for points inside. This property offers a straightforward way to determine whether a point is inside or outside the particle, providing possibilities for a more efficient contact-detection scheme\cite{zhang2021metaball, zhao2022metaball}. It also decides the loss function for the gradient search (GS) in Metaball-Imaging as discussed in Section \ref{MI_step2}. 

\begin{figure}
    \begin{centering}
    \includegraphics[width=0.8\linewidth]{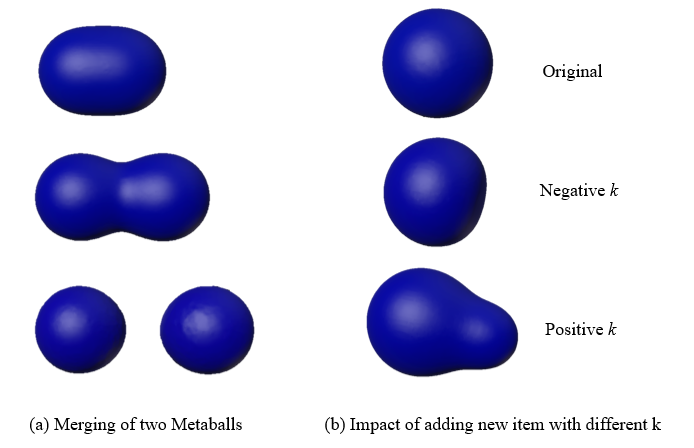}
    \caption{Visualization of 3D Metaballs. (a) stands for two Metaballs getting closer to each other (b) reveals the influence of $\hat{k_i}$ value, where adding a positive Metaball can result in protrusion and a negative Metaball for indentation.}
    \label{fig:visualMetaballs}
    \end{centering}
\end{figure}

Moreover, such function form endows the model a strong physical meaning. This benefits the treatment of translation, rotation and scaling in coupled mechanical models. It also creates convenience for the shape expression. As shown in Figure \ref{fig:visualMetaballs} (a), only a small number of control points(black circles) are required to represent a complex smooth surface. This reveals that the Metaball model can relieve the trade-off between characterization fidelity and model complexity, avoiding artificial surface roughness of the clustering scheme\cite{wang2022novel}. Besides, Figure \ref{fig:visualMetaballs} (b) shows that both protrusion and indentation can be expressed with different settings of $\hat{k_i}$. This indicates that the detailed particle surface texture can be described by adding new control points. Thus, it is possible for the Metaball descriptor to represent highly complex-shaped particles at low cost, even non star-like and concave-void ones. An example is shown in Figure \ref{fig:visual_Image_Avatar}. Related studies also indicate that the Metaball descriptor can be used to represent the deformation and failure processes effectively\cite{plankers2001articulated, jin2000general}. The above advantages together make the Metaball descriptor a proper choice in representing particle morphologies.

\subsection{Metaball Imaging}\label{structure_MI}
\begin{figure}
    \begin{centering}
    \includegraphics[width=1.0\linewidth]{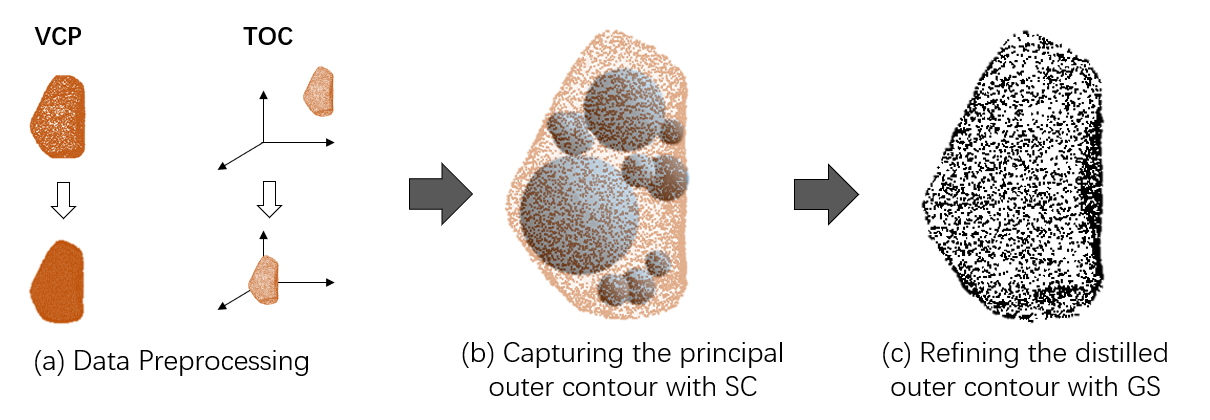}
    \caption{The framework of Metaball Imaging}
    \label{MI_G}
    \end{centering}
\end{figure}

Metaball-Imaging (MI) is developed to transform the XRCT image of irregular-shaped particle into an explicit, Metaball-function based mathematical representation, which is called MI {\it avatar} in this paper. It possesses key necessary morphological information of the given particle for further mechanical simulations. MI is first proposed by authors in previous works\cite{zhao2022metaball}. Here, its fitting ability is greatly enhanced by considering geometric information. The upgraded MI consists of three major parts, data pre-processing, capturing principle outer-contour with sphere-clustering (SC) and refining the distilled contour with gradient-search (GS). The general framework is as shown in Figure \ref{MI_G}.

\subsubsection{Data pre-processing}\label{MI_step0}
Two strategies are adopted in pre-processing(Figure \ref{MI_G}, a), voxelization of cloud-points(VCP) and transformation of coordinates (TOC). VCP is used to voxelize the cloud points of targeted particle, where the voxels are set to be one. TOC is implemented to translate the cloud-points of interested particles into the coordinate system centered at the origin. Such an operation is dedicated to obtain centralized point hull, which can avoid abnormal fitted parameters caused by XRCT coordinates. The distilled voxelized representation and point hull are noted separately as $\boldsymbol{V}$ and $\boldsymbol{H}$ in this paper.

\subsubsection{Capturing principle outer-contour with sphere-clustering}\label{MI_step1}
Here we develop an algorithm to capture the principle outer-contour by searching for a series of non-overlapping inscribed spheres as the control spheres(Figure \ref{MI_G}, b). The flowchart of this algorithm is as shown in Algorithm \ref{alg:sc}. Initially, the Euclidean distance transform\cite{bailey2004efficient} is carried out on the pre-processed voxelated image. Then, the radius $r_i$ and centre $\boldsymbol{c_i}$ of the maximum inscribed sphere can be found with the largest value and its corresponding position in the transformed vector. Next, the voxels belonging to the inscribed sphere are zeroed. The above operations are repeated until satisfying the termination condition. Here, it is defined as the number of inscribed spheres, which is equal to the number of control points $n$. In the end, a set of inscribed spheres $S_I$ can be obtained.

\begin{algorithm}
    \caption{The Sphere-Clustering Algorithm for capture of the principal outer contour}\label{alg:sc}
    
    \hspace*{\algorithmicindent} \textbf{Input:} The voxelized particle $\boldsymbol{V}$, the number of control points $n$
    
    \hspace*{\algorithmicindent} \textbf{Output:} The distilled set of inscribed spheres $\boldsymbol{S}_{I}$
    
    \begin{algorithmic}[1]
    \For{$i = 1,2,...,$ to $n$}
        \State \textbf{Transform -} Implementing Euclidean distance transform on $\boldsymbol{V}$;
        \State \textbf{Search -} Finding the radius $r_i$ and centre $c_i$ of maximum inscribed sphere with the maximum value in the transformed vector;
        \State \textbf{Reset -} Zeroing those voxels of the searched inscribed circles and updating $\boldsymbol{V}$;
    \EndFor
    \State \textbf{Return:} The distilled set of inscribed spheres $\boldsymbol{S}_{I} =$  \{$r_i, \boldsymbol{c_i}$\}, $i$ $\in$ $[1, n]$.
    \end{algorithmic}
\end{algorithm}

\subsubsection{Refining the distill contour with gradient-search}\label{MI_step2}
In this section, the gradient-search (GS) is applied to refine the distilled outer-contour. Since this technique is already presented in \cite{zhao2022metaball}, only key points are discussed here, as listed in Algorithm \ref{alg:GS}.. 
With the distilled set of inscribed spheres $\boldsymbol{S_I}$, a Metaball model $\boldsymbol{M_r}=\{r_i, \boldsymbol{c_i}\}$ can be used to represent the principle outer-contour of the target particles. Then, a loss function is defined to calculate the gradient information provided by the distilled point hull $\boldsymbol{H}$. Here, we utilize a piecewise function instead of the popular Mean Square Error (MSE): 
\begin{equation}
    L(\boldsymbol{M_r} )= \begin{cases}\sum_{i=1}^{m}(f_{\boldsymbol{H_i}}^l(\boldsymbol{M_r})-1)^{2}, & f_{\boldsymbol{H_i}}^l \in[2,+\infty) \\ \sum_{i=1}^{m}(f_{\boldsymbol{H_i}}^l(\boldsymbol{M_r})-1), & f_{\boldsymbol{H_i}}^l \in[1,2] \\ 
    \sum_{i=1}^{m}\left[(f_{\boldsymbol{H_i}}^l(\boldsymbol{M_r})-1)^{2}+\frac{1}{f_{\boldsymbol{H_i}}^l(\boldsymbol{M_r})}-1\right], & f_{\boldsymbol{H_i}}^l \in[0,1]\end{cases}
    \label{GSloss}
\end{equation}
where $m$ is the number of control points and $f_{\boldsymbol{H_i}}^l(\boldsymbol{M_r}) =\sum_{j=1}^{n} \frac{{r_{j}}}{\left(\boldsymbol{H_j}-\boldsymbol{{c}_{j}}\right)^{2}}$.


The reasons for implementing the above function are two-fold. On the one hand, it can avoid obtaining distorted Metaball models. As mentioned in Section \ref{metaball}, when the study point is internally close to or externally far from the Metaball surface, its Metaball function value will all be close to 1. This can make the corresponding MSE very small, resulting in local optimal GS solutions, distorted models with control points outside the targeted surface\cite{zhao2022metaball}. The proposed function effectively remedies this problem, by amplifying the loss value of these points. On the other hand, it can also improve the adaptability to complex geometry, since the amplification of loss values allows GS to capture more gradient information, especially on the surface texture. 

In the end, the model parameters are optimized through gradient descent:
\begin{equation}
    \boldsymbol{M_r} \leftarrow \boldsymbol{M_r} -\eta \cdot \nabla_{\boldsymbol{M_r} } L(\boldsymbol{M_r} )
\end{equation}
where $\eta$ is the learning rate; $\nabla_{\boldsymbol{M_r}} L(\boldsymbol{M_r} )$ is the gradient of the $L(\boldsymbol{M_r} )$ to the parameter $\boldsymbol{M_r}$. As for gradient update, Adaptive Moment Estimation (Adam) and Stochastic Gradient Descent (SGD) \cite{zhang2018improved} are combined to achieve a balance in computational efficiency and convergence ability\cite{wang2018optimization}. The above function is repeated until satisfying the termination condition, which is the number of generations $E^{gs}$.

\begin{algorithm}
    \caption{The Gradient Search for refinement of outer contour}\label{alg:GS}
    
    \hspace*{\algorithmicindent} \textbf{Input:} the particle point hull $\boldsymbol{H}$, the number of generations $E^{gs}$, the learning rate $\eta$, the distilled set of inscribed spheres $\boldsymbol{S_I}$.\\   
    \hspace*{\algorithmicindent} \textbf{Output:} the metaball model of the refined outer contour $\boldsymbol{M}_{f}$.
    
    \begin{algorithmic}[1]
    \State $\boldsymbol{S}_{I}$ is taken as the Metaball model of principle outer-contour, the initial value $\boldsymbol{M_r}$; 
    \For{$i = 1,2,...,$ to $E^{gs}$} 
        \State $\boldsymbol{M_r} \leftarrow \boldsymbol{M_r}  -\eta \cdot \nabla_{\boldsymbol{M_r}} L(\boldsymbol{M_r})$;
    \EndFor

    \State \textbf{Return:} The searched parameter $\boldsymbol{M}_{f}$.
    \end{algorithmic}
\end{algorithm}

\section{Evaluation of Metaball-Imaging }\label{validation_MI}
In this section, we discuss the characterization performance of the proposed MI by examining four types of grains with different geometric characteristics. 

\subsection{Dataset and XRCT setting}\label{MI_exp}
\begin{figure}
    \centering
   \includegraphics[width=1.0\linewidth]{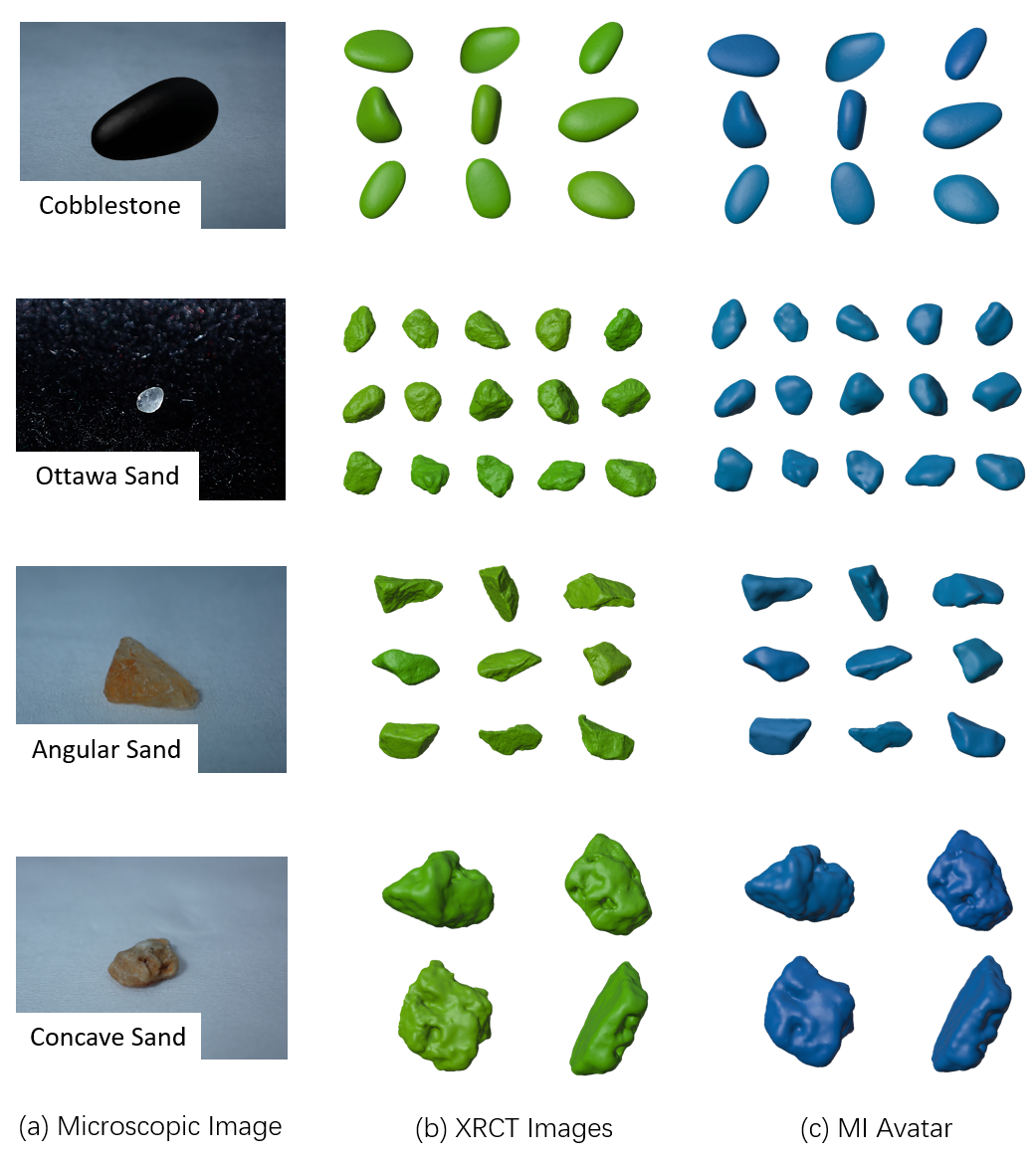}
    \caption{Representative particles, XRCT images and corresponding MI avatars of the selected four types of particles. 
    Numbers of control points to reconstruct the cobblestone, Ottawa sand, angular sand and concave sand are 40, 40, 100 and 120 respectively.}
    \label{fig:visual_Image_Avatar}
\end{figure}

Four types of particles, which are popular in the research field, are chosen for evaluation. They are 20 cobblestones, 290 Ottawa sands, 25 angular sands and 5 concave sands, as shown in Figure \ref{fig:visual_Image_Avatar} (a). The cobblestone is a common granite or basalt in the riverbed. It possesses obvious round features, which are the results of water flow. The Ottawa sand is a typical grain of pure quartz exploited from Ottawa in Canada. Due to geological transportation, it features smooth and angled characteristics. As for the angular sand and the concave sand, they are sand for common building materials mainly composed of quartz and feldspar. They have angular and concave characteristics caused by mechanical disruption and chemical weathering.

In XRCT imaging, the ZEISS Xradia 610 Versa is utilized. The voltage of X-ray source is set to 140kV for cobblestone and 80kV for the others. The 0.4X detector is chosen in the scan recipes, which means the corresponding optical magnification is 0.4. The voxel size is 45.48$\mu m$ for the cobblestone, 18.56$\mu m$ for the Ottawa sand, 51.64$\mu m$ for the angular sand, 19.59$\mu m$ for the concave sand. On average, the particle in XRCT images contains more than $7.9 \times 10^6$ voxels to represent a real grain geometry. The particle segmentation is done by "ilastik", a machine-learning driven edge-detection algorithm for XRCT images \cite{sommer2011ilastik}. The "ilastik" algorithm can detect particle edges by considering both voxel intensity and brightness gradient, which effectively avoids omitting particles with concave spots on the surface or broken spots on the edge. The processed particles are shown in Figure\ref{fig:visual_Image_Avatar} (b). 

\subsection{Metrics}
We select seven shape factors for evaluation: surface area $A$, volume $V$, Corey Shape Factor $CSF$, nominal diameter $D_n$, surface-equivalent-sphere diameter $D_s$, sphericity $\phi$ and circularity $C$.

The Corey Shape Factor $CSF$\cite{dietrich1982settling} reveals the dimension feature of the studied particle, as given by:
\begin{equation}
    \mathrm{CSF}=\frac{L_{s}}{\sqrt{L_{i} L_{l}}}
\end{equation}
where $L_s$, $L_i$ and $L_l$ are the shortest, intermediate and longest axis lengths of particles. 

The nominal diameter $D_n$ and surface-equivalent-sphere diameter $D_s$ are two widely used parameters\cite{bouwman2004shape, zhang2016lattice}. The $D_n$ is defined as the diameter of the volume-equivalent sphere. And the $D_s$ takes the following form:
\begin{equation}
    D_{s}=\sqrt{\frac{4 A_{p}}{\pi}}
\end{equation}
where $A_{p}$ = the maximum projected area of the particle. Here, they are combined as $D_{ns} = D_n / D_s$ to form a dimensionless quantity. 

The sphericity $\phi$ \cite{mora2000sphericity} is the measure of similarity between the studied particle and the sphere, which is defined as:
\begin{equation}
    \phi=\frac{A_{ve}}{A}
\end{equation}
where $A_{ve}$ = the surface area of the volume-equivalent sphere to the studied particle; $A$ = the surface area of the studied particle. 

Another frequently used metric is the circularity $C$ \cite{bouwman2004shape}, which evaluates the roundness of non-spherical particle:
\begin{equation}
    C=\frac{\pi D_s}{P_p}
\end{equation}
where $P_p$ is the the perimeter of the particle's projected-area.

\subsection{Impact of the control-point number on characterization quality}
In Metaball-Imaging, a vital variable is the number of control points $n$, which is closely related to fidelity. For applications, it is necessary to understand the relationship between control point number and characterization quality. Evaluations are carried out on multiple, randomly-selected particles of each type.   

\begin{figure}
    \centering
    \includegraphics[width=0.8\linewidth]{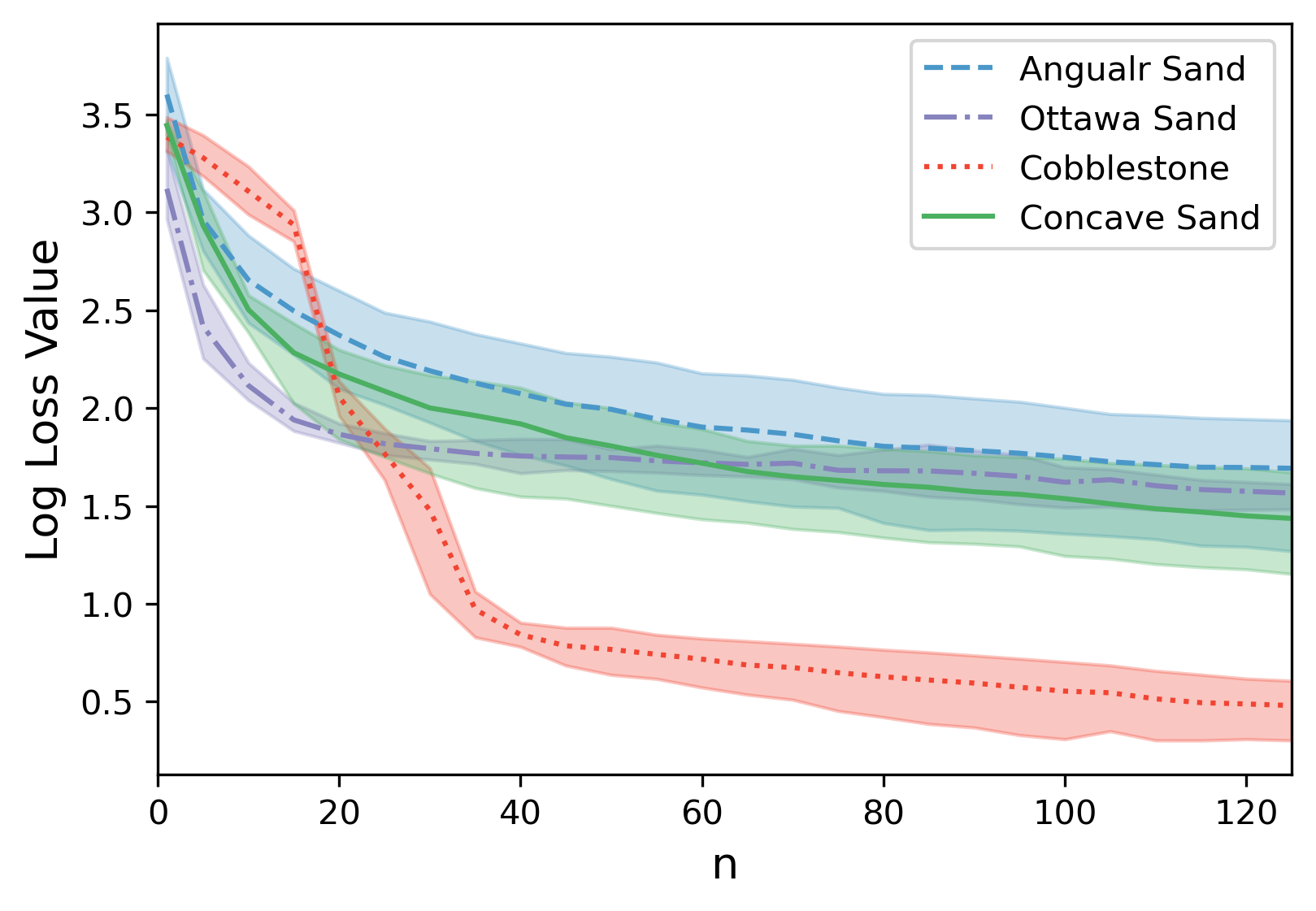}
    \caption{Correlations of the loss value with the number of control points for four types of particles, where the log loss value is base-10 logarithm of the final value of Eq. \ref{GSloss} in gradient search. }
    \label{fig:n2fidelity}
\end{figure}

The converged value of the loss function (Eq. \ref{GSloss}) in GS is a critical index for characterization. As shown in Figure \ref{fig:n2fidelity}, the log loss values of studies particles decrease rapidly as the increase of $n$. When $n$ equal to 40, all of them reached valleys, where further increasing the control points will not decrease the loss function significantly. Compared with other particles, the cobblestone has a very different loss curve. Its loss value first decreases slowly, where limited control points are hard to represent cobblestones with flat features. Then, the loss value starts to drop rapidly as more control points are involved until converges at $n$ equal 40, following a similar pattern of the other particles. Note that it takes much less control points (about 20) for the cobblestone to reach the converged loss value of the rest particles and its converged value is also lower. These together indicate the superiority of Metaball function in representing particles with smooth, round features.

Figure \ref{fig:n2rec} illustrates the morphology of typical particles reconstructed by MI with different settings of $n$. It is shown that the increasing control points allow MI avatar to capture more morphological details. The reconstruction fidelity first improves with the increase of control points and then become steady, which indicates a proper $n$ can avoid too many avatar parameters while guaranteeing reconstructed fidelity. It is worth noting that the number of control points required for different types of particles is different. For the cobblestone and Ottawa sand, dozens of control points are enough for characterization. In contrast, more points are needed for the capture of angular and concave sand, which possess more complex geometrical features.  

\begin{figure}
    \centering
    \includegraphics[width=0.9\linewidth]{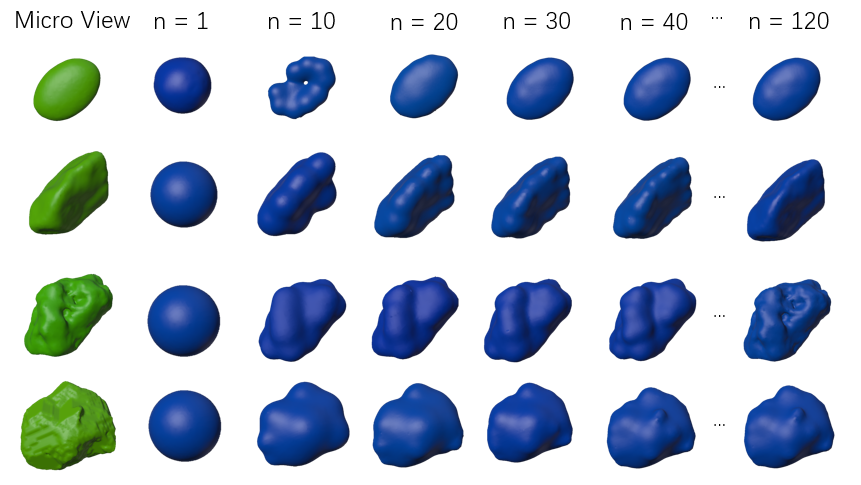}
    \caption{Impact of the control point number on the reconstructed particle morphologies }
    \label{fig:n2rec}
\end{figure}

To select $n$ properly, we carry on further evaluations on more precise geometric metrics of typical particles (See Figure 7). The relative error $\delta$ between the real particle and MI avatar is selected for comparison.
It is found that all $\delta$ of chosen metrics converged rapidly to low stable values after $n$ greater than 40, which indicates a good match between parents and MI avatars. However, the characterization quality of each indicator varies considerably. The sphericity and circularity are much easier to reconstruct compared to the rest. And small deviations can be observed in some metrics, like volume and surface area of angular and concave sands. 

Such phenomena are mainly attributed to those minor details of the XRCT images, such as the surface roughness and concave holes. These details will interfere the imaging quality of real particles. Theoretically, a better agreement can be achieved with MI avatars of more control points. In fact, those error curves are not converged yet and are still slowly decreasing. However, a refined MI avatar may not be idealized in application due to possible errors in image measured value of the real particle. 
Taking an example on the image measured result of surface area, the source of error can either be from the selection of XRCT optical magnification or treatment of voxel connectivity, which might make the scanned "real" particle possess limited fidelity and show unrealistic wrinkle characteristics. On the contrary, the MI avatar is distilled directly from the cloud points, of which the results are more convincing. 
A similar problem is also reported in \cite{zhou2015micromorphology}, where a 10$\%$ relative error is observed on the image-measured surface area compared to the reconstructed SH value. 

In conclusion, the MI avatar with n of 40 is found to be sufficient to reconstruct the major morphological features and surface texture for all four types of particles involved in the current study. For higher fidelity characterization of angular and concave features, a MI avatar with more control points, offering more geometric details, can be a better choice.

\begin{figure}
    \centering
    \renewcommand{\thefigure}{\arabic{figure} I}
    \setcounter{figure}{6}
   \includegraphics[width=1.0\linewidth]{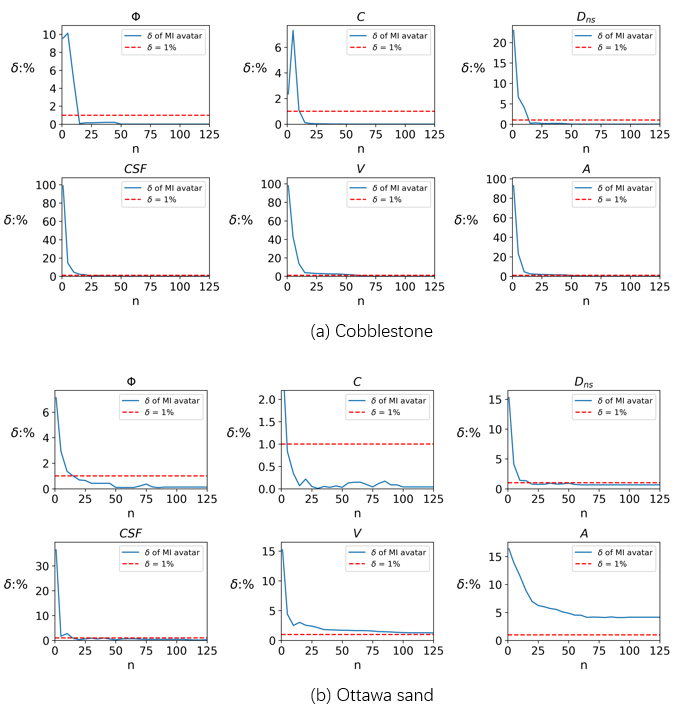}
    \caption{Impact of the control point number on reconstructed geometric metrics}
    \label{fig:n2rec_cobble_sand}
\end{figure}

\begin{figure}
    \centering
    \renewcommand{\thefigure}{\arabic{figure} II}
    \setcounter{figure}{6}
   \includegraphics[width=1.0\linewidth]{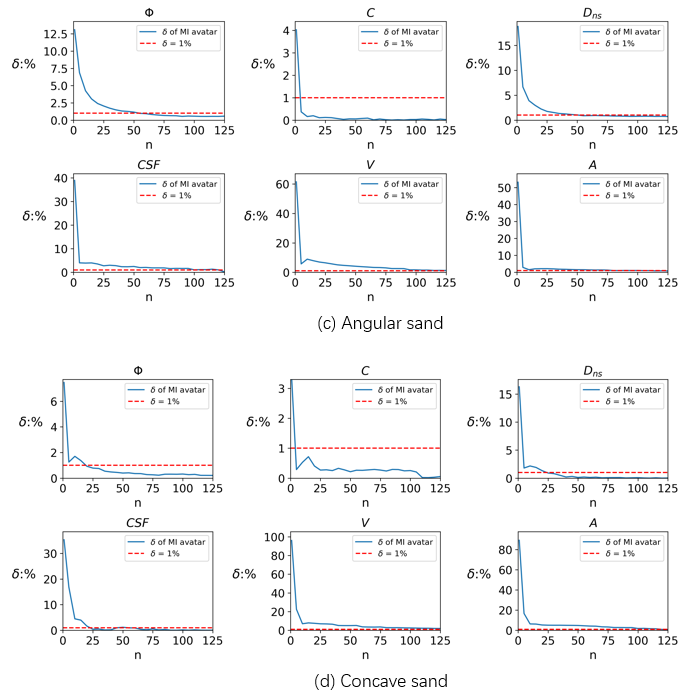}
    \caption{Impact of the control point number on reconstructed geometric metrics}
    \label{fig:n2rec_angular_concave}
\end{figure}





\subsection{Reconstruction performance on particle groups}
To further evaluate the performance of MI, experiments are implemented on particle groups introduced in Section \ref{MI_exp}. MI avatars of 40 control points are adopted to reconstruct the cobblestone and Ottawa sand. As for angular sand and concave sand, 100 and 120 control points are applied separately. The learning rate of gradient search is chosen to be 0.001. 

Visualizations of randomly-selected parental particles and MI avatars of each type can be found in (b) and (c) of Figure \ref{fig:visual_Image_Avatar}. Such a good match can also be observed in cumulative distribution functions (CDFs) of shape indicators as shown in Figure 8. For example, Figure 8 (b) compares the CDF of the six shape factors between the parents and MI avatars of the Ottawa sand. A good agreement is found. Noting that small deviations can be found on some metrics of the rest particle groups. This is mainly caused by the limited particle number rather than the reconstruction error.

\begin{figure}
    \centering
    \renewcommand{\thefigure}{\arabic{figure} I}
    \setcounter{figure}{7}
    \includegraphics[width=1.0\linewidth]{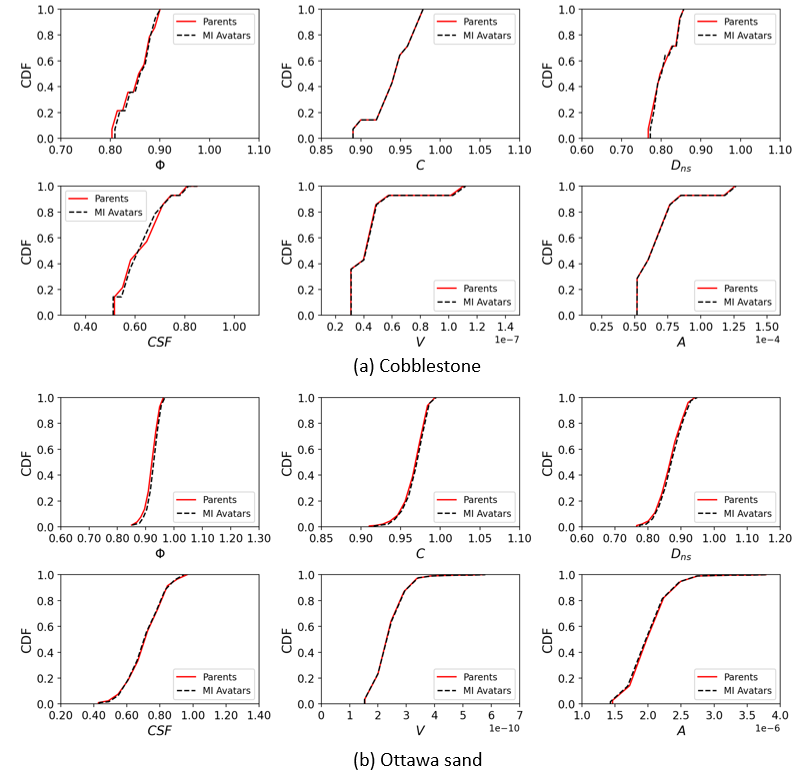}
    \caption{Reconstruction quality of morphological metrics on particle groups}
    \label{fig:rec_cobble_sand}
\end{figure}

\begin{figure}
    \centering
    \renewcommand{\thefigure}{\arabic{figure} II}
    \setcounter{figure}{7}
    \includegraphics[width=1.0\linewidth]{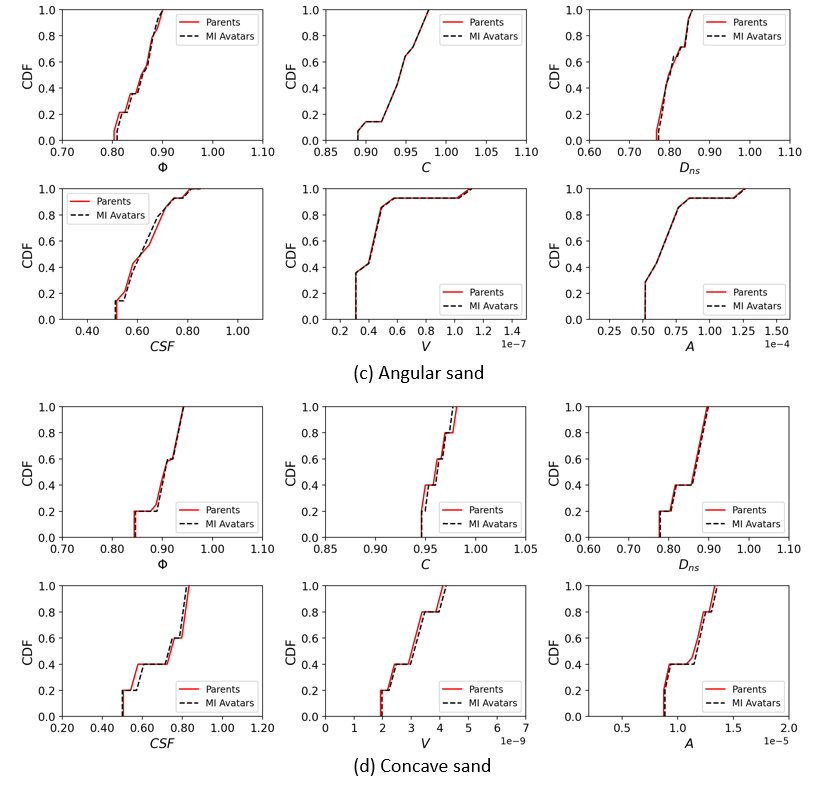}
    \caption{Reconstruction quality of morphological metrics on particle groups}
    \label{fig:rec_angular_concave}
\end{figure}

The above results are calculated with Python in the parallel mode on a personal workstation (Technical details listed in Table \ref{pcSetting}). A total of 340 avatars are characterized from their XRCT images and the average time taken to characterize each particle is 1.03 minutes.

\begin{table}\centering
\begin{tabular}{cclcl}
\cline{1-4}
Processor            &                      &  & Intel Xeon W-2133 3.60GHz &  \\
RAM                  &                      &  & 32GB  2400MHz             &  \\
\cline{1-4}
\multicolumn{1}{l}{} & \multicolumn{1}{l}{} &  & \multicolumn{1}{l}{}      & 
\end{tabular}\label{pcSetting}
\end{table}

\section{Generating 3D Style-similar Granular Particles in Identically Distributed Morphological Features with Metaball Descriptor}\label{MetaballVAE}
\subsection{MetaballVAE}\label{structure_MetaballVAE}






The variational autoencoder(VAE)\cite{kingma2013auto} is a neural-network based generative model. The framework of it is to first learn the distribution $P$ of the targeted data $\boldsymbol{x}$ and then generate through sampling with some unobserved variable $\boldsymbol{z}$, which is called the latent variable. And the collection of them is named the latent space. In implementation, the learning of $P(\boldsymbol{x})$ is carried out with an assumed distribution $\int Q(\boldsymbol{z}\mid \boldsymbol{x})Q(\boldsymbol{x}) d\boldsymbol{z}$ ($Q$ is the assumed distribution of two parts. Since these two parts are implemented in one neural network system, they share the same notation)
, which is in the form of neural network. This distribution corresponds to two important components of VAE: the encoder $\boldsymbol{z}=E(\boldsymbol{x})$(For $Q(\boldsymbol{z} \mid \boldsymbol{x})$) and decoder $\bar{\boldsymbol{x}}=D(\boldsymbol{z})$(For $Q(\boldsymbol{x})$), where $\boldsymbol{x}$ represents the input, $\bar{\boldsymbol{x}}$ for the generated(reconstructed). For particle generation, $\boldsymbol{x}$ and $\bar{\boldsymbol{x}}$ refer to the shape representation, for example, the Metaball descriptor $\boldsymbol{M}$ or XRCT images. The encoder and decoder consist of the major steps in VAE: encoding and decoding.  In encoding, the input shape representation $\boldsymbol{x}$ is compressed and mapped into the latent variable $\boldsymbol{z}$, a multidimensional shape-representation tensor. Then $\boldsymbol{z}$ is decoded to reconstruct the input particle $\bar{\boldsymbol{x}}$. Through minimizing the difference between $\boldsymbol{x}$ and $\bar{\boldsymbol{x}}$, morphologies and shape-feature distribution of input particles can be learned effectively (The theory behind this is briefly stated in Appendix A). Then, the trained decoder $\bar{\boldsymbol{x}}=D(\boldsymbol{z})$ can be applied to generate particles by inputting random $\boldsymbol{z}$. Assisted by the powerful learning ability of neural network, VAE can inference new particles, which are not included in the training set but maintain coessential morphological features and distributions with the parental particles. 

Previous studies on particle generation follow a similar scheme\cite{zhou2015micromorphology, zhou2017generation, wei2018generation, xiong2021gene}. They usually compress morphological patterns of particle representation $\boldsymbol{x}$ (e.g. XRCT images) into some controlling variables (e.g. SH descriptor), similar to latent variables $\boldsymbol{z}$ in VAE. Then, various algorithms are implemented to add small randomness into this controlling variable to generate new particles. However, direct sampling on those controlling variables can result in underfitting or overffiting problems in shape-feature distributions of generated particles\cite{wei2018generation, shi2021randomly}. This can be relieved with mixture models\cite{medina2019geometry, xiong2021gene}. But the shape feature of realistic particles have a large variance range and the corresponding feature distributions could be too complex to be manipulated with explicit methods\cite{shi2021randomly, macedo2023shape}. With deep neural-network system, VAE can provide a more practical solution. Note that latent variables $\boldsymbol{z}$ in VAE are regularized (chosen to be multivariable normal distribution) to encourage similar input samples compress at closer positions in the latent space. This property allows the model to learn a more flexible and general distribution, rather than simply adapting to the specific patterns present in the training data. As a result, generation by sampling from sperate, regularized $\boldsymbol{z}$ can help to avoid overfitting and underfitting\cite{shi2021randomly}. More importantly, it enables high-level controls on the generated particle morphologies\cite{macedo2023shape}.

Based on VAE, we propose a Metaball-based particle generation framework, called Metaball Variational Autoencoder(MetaballVAE). It can resolve the correlation between XRCT images and morphologies of input particles with a regularized latent space, where complex calculations are converted into one-step solutions, reducing variance in generated models and improving controls on generation process. This model requires no prior knowledge (e.g. particle shape-feature distributions) but only XRCT images of the target particles to generate non-existent style-similar avatars, particles in the form of Metaball, which can be put into simulations directly. Note that the style-learning is not limited on the morphology but also the shape-feature distribution, including surface area, volume, sphericity etc.

\begin{figure}
    \centering
    \includegraphics[width=1.0\linewidth]{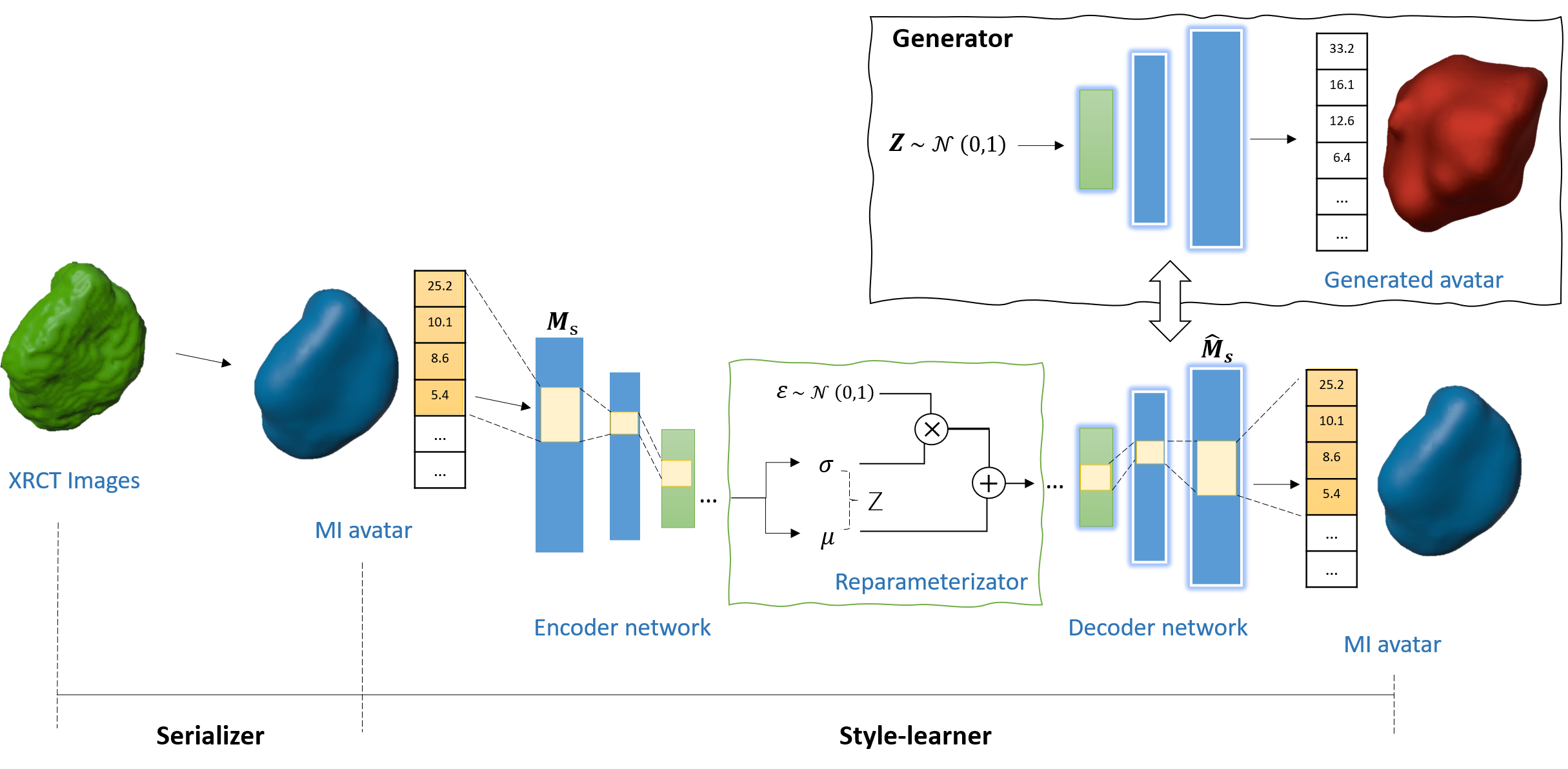}
    \caption{MetaballVAE for generation of complex-shaped particles. Serilizer: interpreting and transforming XRCT images of granular matters. Style-learner: analysing the serilized Metaball descriptor and learning the morphological characteristics and distribution. Generator: generating style-similar granular media in Metaball form}
    \label{fig:metaballVAE_showup}
\end{figure}

The MetaballVAE consists of three major parts as illustrated in Figure \ref{fig:metaballVAE_showup}: serializer, style-learner and generator. The serializer interprets and transforms XRCT images of target granular-particles into Metaball descriptors. Then, the style-leaner analyses those distilled descriptors, capture major shape characteristics, conducts inference on feature distribution and devise style-similiar avatars. In the end, the generator outputs designed style-similar, Metaball-based avatars. 

\subsection{The serializer}  
The serializer is designed to abstract the particle morphology, extract shape-feature distributions and code them structurally. It can significantly reduce the dimension of XRCT images while keeping all the vital morphological information for generation. The reasons for implementing it are two-fold. On the one hand, structured data are more suitable for generation tasks\cite{para2021sketchgen}, which can improve the generation quality. On the other hand,  this enables a direct generation of particles in Metaball function form, which can be put into simulations without bridging or transformation, avoiding unnecessary information loss and computational cost.


In this paper, the serialization is accomplished with the Metaball-Imaging technique as introduced in Section \ref{structure_MI}. The serilized particle is in MI avatar form, which is noted as $\boldsymbol{M_s}$. 

\subsection{The style-learner}


The style-learner is a modification of the aforementioned VAE. It can digest the structured data $S$ and learn how to generate particles. Main components involved are: encoder, decoder, reparameterizator, loss function and distribution annealer. 

\textbf{Encoder} and \textbf{Decoder} are multi-layer perceptions, which are connected in a bottle-neck form as shown in Figure \ref{fig:metaballVAE_showup}. They are implemented to approximate the real distribution $P(\boldsymbol{x})$ as a learnable, assumed distribution $\int Q(\boldsymbol{z} \mid \boldsymbol{x})Q(\boldsymbol{x})dz$. On the topic of particle generation, $\boldsymbol{x}$ refers to the $\boldsymbol{M_s}$. The encoder takes serialized particles $\boldsymbol{M_s}$ as input and outputs parameters($\mu$ - the mean, $\sigma$ - the standard deviation) of the corresponding latent variable $\boldsymbol{z}$, mapping morphologies and shape-feature distributions of particles into a regularized latent space. On the contrary, the decoder interprets $\boldsymbol{z}$ to restore $\bar{\boldsymbol{M_s}}$, reconstructing particle morphologies and shape-feature distributions from that regularized space.  

\begin{figure}
    \centering
    \includegraphics[width=1.0\linewidth]{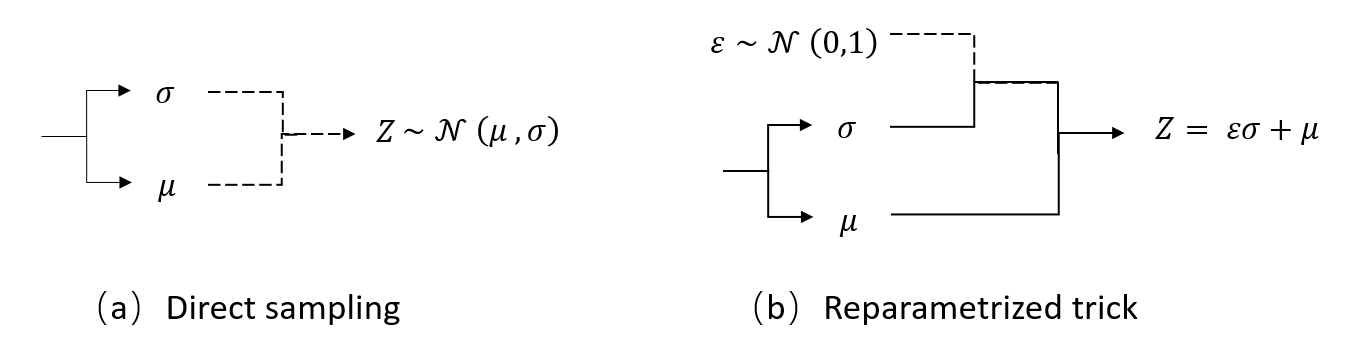}
    \caption{Visualizations of the direct sampling and reparameterization trick. The solid line stands for relationship capable of back propagation.The dash line represent relationship where backpropagation can not be carried out. }
    \label{fig:repara}
\end{figure}

\textbf{Reparameterizator} locates halfway between the encoder and decoder. It is designed to regularize the latent space by creating a map between the encoded information and a normal distribution. Instead of direct sampling (Fig, \ref{fig:repara}, a), a more deterministic pattern is utilized: 

\begin{equation}
z=\mu+\sigma \odot \epsilon    
\end{equation}
where the $\epsilon$ is the assumed normal distribution. This enables continuous gradient calculation on the mapping relationship, making MetaballVAE a learnable system. 

\textbf{The loss function} is the global optimization objective for MetaballVAE:

\begin{equation}
L(\boldsymbol{M_s}) =  \underbrace{\frac{1}{d} \sum_{k=1}^d {\|\boldsymbol{M_s} - \hat{\boldsymbol{M_s}}\|}^2}_{\text {Reconstruction Item}}  + \underbrace{\frac{1}{2} \sum_{k=1}^d\left(\mu_{(k)}^2(\boldsymbol{M_s})+\sigma_{(k)}^2(\boldsymbol{M_s})-\ln \sigma_{(k)}^2(\boldsymbol{M_s})-1\right)}_{\text {Distribution Item}} 
\label{loss_function}
\end{equation}
where $d$ is the dimension of $\boldsymbol{M_s}$. This function is modified from the original VAE theory based on the particle generation problem. The deduction of it is stated in Appendix B. It consists of two items: the distribution item and reconstruction item. The distribution Item measures the difference between real and learned distributions of particle morphologies. The Reconstruction Item evaluates the quality of learned morphological characteristics. The combination of them forces MetaballVAE to learn not only morphologies but also shape-feature distributions of the input particles. 

\textbf{The distribution annealer} is proposed to tackle the training challenge of VAE. A well-trained model possesses a relatively small reconstruction item and a non-zero distribution item. However, most direct training will yield a model with a zero distribution item. Such tendency in learning is caused by the sensitivity of decoder to variation introduced by the mapping process of reparameterizator. This makes the decoder ignore the latent variable provided by the encoder and output the average optimal with distribution item equal to zero. For this reason, the distribution annealer is implemented by adding a weight to the distribution item of the loss function(Eq. \ref{loss_function}). This weight starts from zero, where the weighted loss function equals the reconstruction loss. Then, the weight is increased gradually to one, where this weighted function satisfies the true loss function definition. With such a process, the model will be forced to use the learned latent space to achieve good likelihood in prediction.  

\begin{figure}
    \centering
    \includegraphics[width=1.\linewidth]{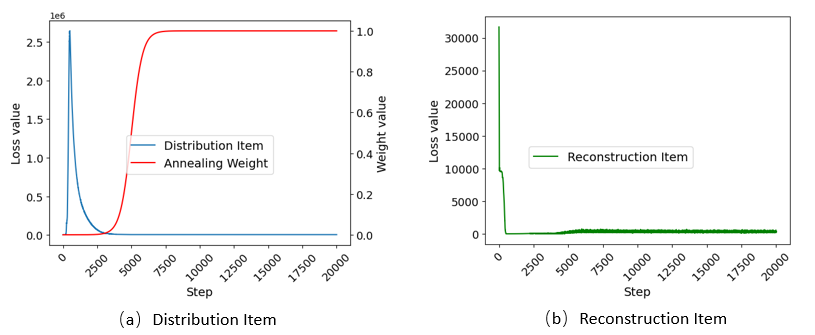}
    \caption{The impact of the distribution annealer on the loss value of different items in the training of cobblestone dataset}
    \label{dis_annealer}
\end{figure}

Figure \ref{dis_annealer} is an example of the distribution annealer in the training of the cobblesteone dataset during the first 20k steps. It can be observed that the distribution item first spikes as the reconstruction item drops significantly, where the model is encoding shape features into the latent space cheaply. Then, the distribution item starts to decrease rapidly as more attention is paid to the divergence penalty. Correspondingly, the decrease of reconstruction item slows down. Finally, the distribution item gradually converges and the reconstruction item enters fluctuation, where more morphological information is compressed into the model.

\subsection{The generator}
Before formal generation, the decoder of style-learner should be well-trained. The generation task requires the trained decoder and a normal distribution $\mathop{N}(0, 1)$, which represents the regularized latent space. A typical generation process is illustrated in Figure \ref{fig:metaballVAE_showup}. The latent variable $\boldsymbol{z}$ is sampled from a normal distribution, serving as the input matrix to the decoder. Then, the decoder can devise style-similar particles unseen in the training dataset. The distribution of shape features can be well reconstructed when the number of generated particles is large enough. It is worth noting that the generated particle is in the form of Metaball descriptor, which can be applied directly into simulations. 


\section{Evaluation of MetaballVAE}\label{Validation_MetaballVAE}

\subsection{Dataset and Metrics}
Previous studies on particle generation are often carried out on hundreds of thousands of samples\cite{shi2021randomly, medina2019geometry, macedo2023shape}. However, particle reconstruction with XRCT requires considerable time and computational resources. In actual engineering, it is very often to have only a dozen scanned particles. It is necessary to test the algorithm performance on smaller datasets. Thus, we evaluate MetaballVAE on two sets of XRCT data with different sizes: 290 Ottawa sands and 20 cobblestones. Details of these two datasets are listed in Section \ref{MI_exp}. For better learning performance, data augmentation are implemented on training datasets, where slightly modified synthetic data is introduced based on the real one. Here, particle rotating and parameter shuffling are implemented. Particle rotating is a popular strategy based on rotational-invariant property. For example, Shi et al. \cite{shi2021randomly} applied nine rotations to each particle and enlarged his dataset by ten times in a particle generation task, which effectively enhanced the model performance. Parameter shuffling means random recombination of $\{k_i, \boldsymbol{x_i}\}$ in the serialized particle $\boldsymbol{M_s}$. This is because the sequence change of control spheres will not modify the corresponding Metaball model. Such processing can effectively avoid the overfitting problem and enhance convergence performance. During augmentation, each particle is rotated 5 times and the corresponding Metaball parameter shuffled 50 times. The augmented datasets then each contain 145,000 Ottawa sand and 6,000 cobblestone samples.  

Accurate evaluation is a challenge for particle generation tasks. Apart from the rationality of particle shape, another important content of evaluation is the quantitative difference between parents and clones. Thus, the same metrics used in characterization are also adopted here for further verification on the shape feature distributions.  

\subsection{The setting of serializer and style-learner}
In this evaluation, we apply the following hyper-parameter setting. In serializer, the control point number, n, is set to be 40 for both cobblestone and Ottawa sand samples. The learning rate for the gradient update is set to be 0.001. In style-learner, the encoder is a 4 layer full-connected network with leaky ReLU activation function. The size of it is: 160$\times$1024$\times$512$\times$256$\times$128. The decoder is also a 4 layer full-connected network with leaky ReLU activation function yet in reverse form. The reparameterizator is set to be one fully connected layer with size 128. The above networks are trained by Adam\cite{zhang2018improved, wang2018optimization} with learning rate $\eta$ = 0.0001.

Since the setting of hyper-parameters is not a focus of this paper, how to obtain them is not included here for the sake of brevity. A detailed procedure can be referenced in \cite{shi2021randomly}.

\subsection{Particle Generation}
In validation, the training datasets are denoted as "Parents". And 1000 particle avatar are devised by the generator, denoted as "Clones". The number of generated particles is set to larger for better evaluation on the morphological distribution. 
 
\begin{figure}
    \centering
    \includegraphics[width=0.7\linewidth]{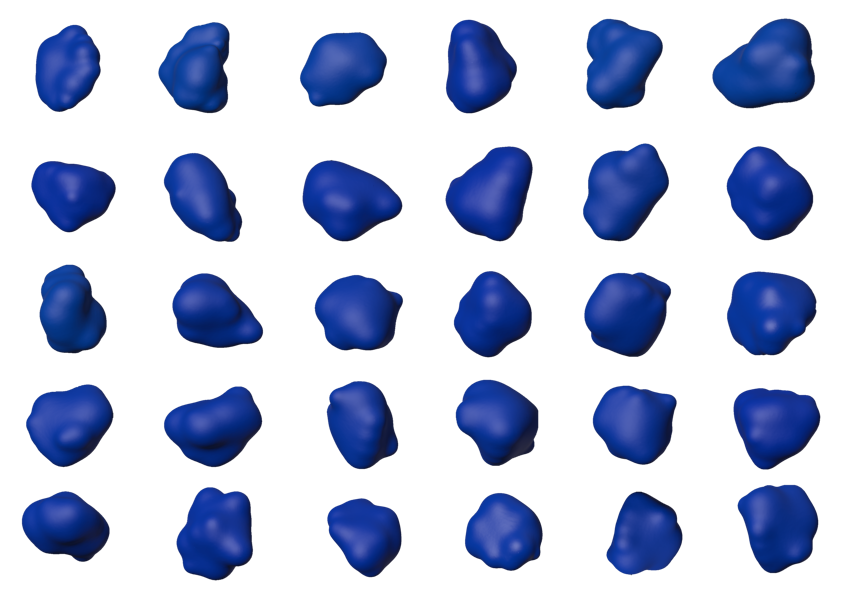}
    \caption{Examples of cloned Ottawa sands}
    \label{fig:visual_generated_sand}
\end{figure}

\begin{figure}
    \centering
    \includegraphics[width=0.7\linewidth]{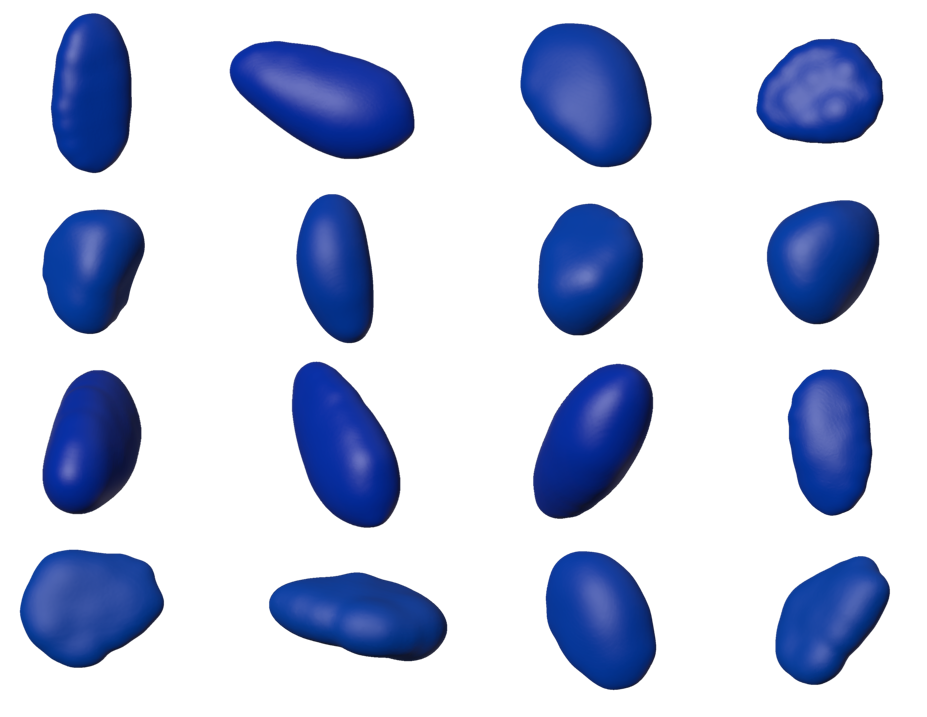}
    \caption{Examples of cloned cobblestone}
    \label{fig:visual_generated_cobble}
\end{figure}

Figure \ref{fig:visual_generated_sand} and Figure \ref{fig:visual_generated_cobble} displays several cloned examples of each type. It can be concluded that these clones exhibit reasonable shape features of both Ottawa sands (angled features) and cobblestones (round features). Note that these particles are in Metaball function form and the meshes are only for visualization. 

\begin{figure}
    \centering
    \includegraphics[width=1.0\linewidth]{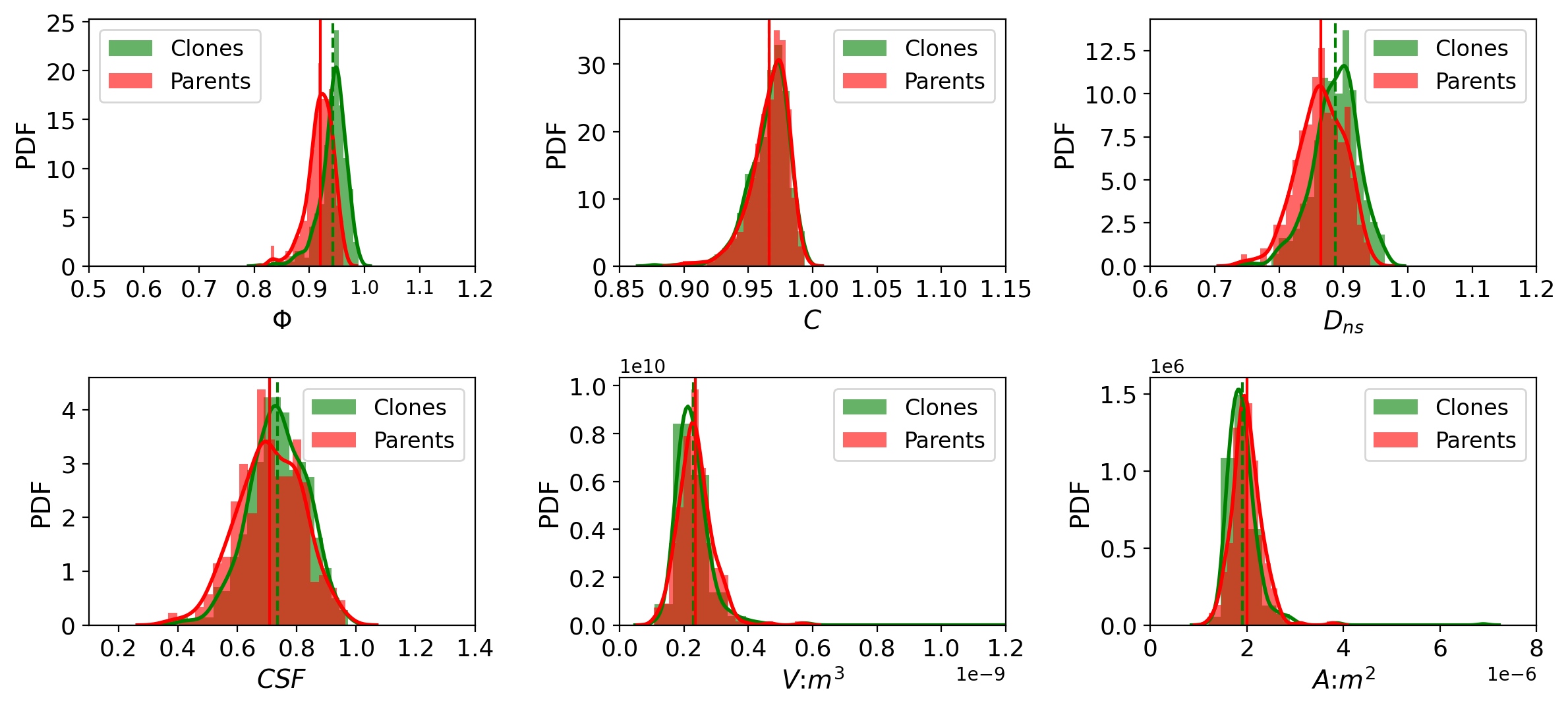}
    \caption{Comparison of feature distributions between parental and cloned particles for the Ottawa sand }
    \label{fig:sand_generation}
\end{figure}

\begin{figure}
    \centering
    \includegraphics[width=1.0\linewidth]{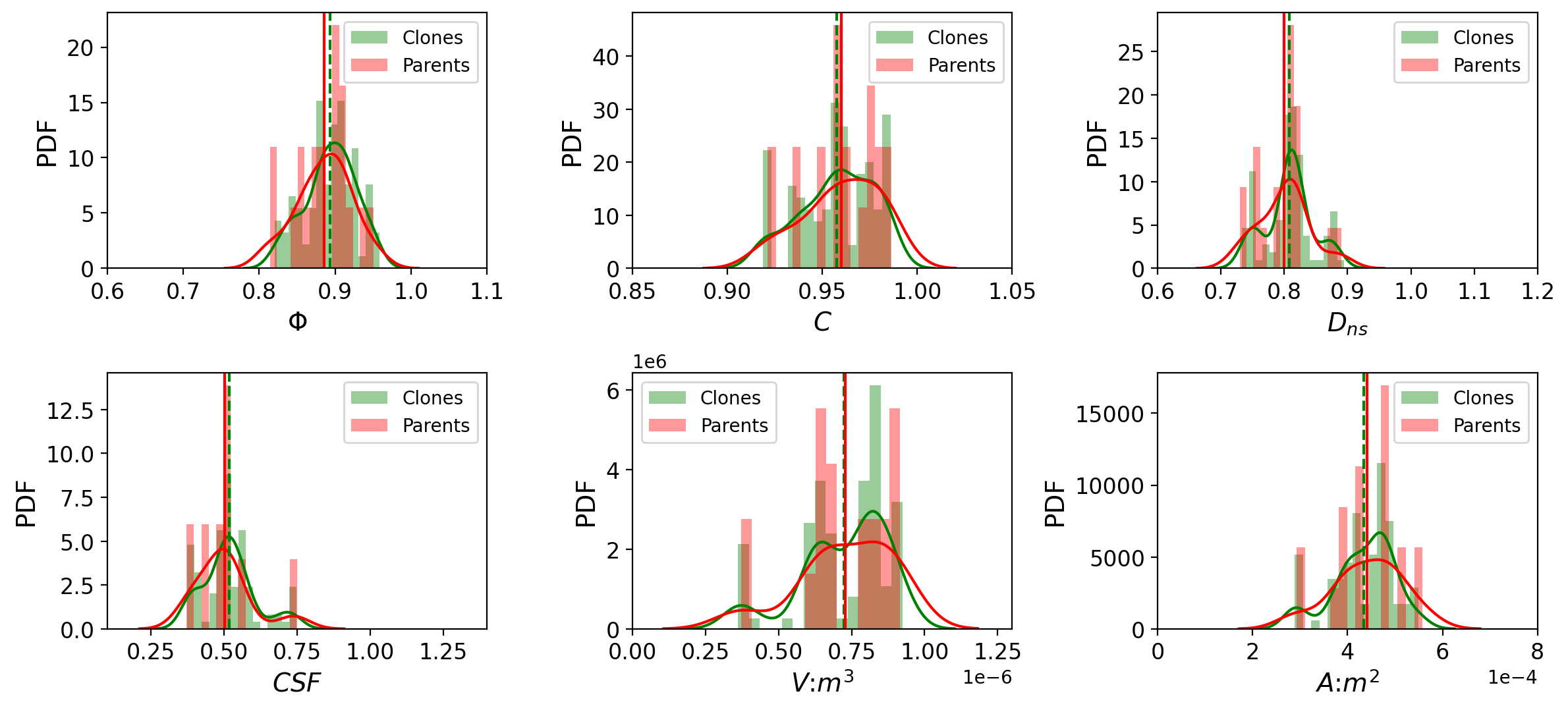}
    \caption{Comparison of feature distributions between parental and cloned particles for the cobblestone}
    \label{fig:cobble_generation}
\end{figure}

To further exam the regeneration ability of the proposed method on shape-feature distributions, probability density functions (PDF) of selected metrics are calculated on both parents and their clones (See Figure \ref{fig:sand_generation} and Figure \ref{fig:cobble_generation}). 

All shape features of parents and clones in both Ottawa-sand and cobblestone datasets share similar distributions. In term of $\phi$, we observe deviation errors of 2.54$\%$ and 0.92$\%$ on means between parents and clones, for the Ottawa sand and cobblestone separately. In the case of standard deviation, errors of 3.12$\%$ and 4.45$\%$ are obtained. For $C$ of the Ottawa sand and cobblestone, the distribution means are off with errors of 0.03$\%$ and 0.25$\%$. While errors in standard deviation are 2.56$\%$ and 3.46$\%$ separately. In the case of $D_{ns}$, errors of 2.57$\%$ and 0.98$\%$ are observed on the mean misalignments for the Ottawa sand and cobblestone. And the errors coming from the standard deviation are 2.69$\%$ and 5.87$\%$. As for $CSF$, the distributions have errors of 3.87$\%$ and 3.25$\%$ on the mean, as well as errors of 11.58$\%$ and 4.88$\%$ on the standard deviations, for the Ottawa sand and cobblestone separately. Finally, we obtain errors of 2.77$\%$ and 0.66$\%$ from $V$, as well as errors of 4.22$\%$ and 1.41$\%$ from $A$ on means of distributions of the Ottawa sand and cobblestone. In the case of standard deviation, the errors are 20.61$\%$ and 5.64$\%$ given by $V$, as well as 24.21$\%$ and 10.65$\%$ given by $A$. 

Note that the MetaballVAE is capable of learning and representing non-Gaussian shape-feature distributions. An example of this can be seen in the histogram of the feature $V$ of the cobblestone, where multiple peaks are observed in the distribution (although they may not be clear in the PDF curve due to limited samples). The MetaballVAE is able to effectively capture this feature and generate particles with similar distributions, demonstrates the effectiveness of the MetaballVAE in cloning grains of complex shape-feature distributions.


Compared with previous studies \cite{zhou2017generation, jerves2017geometry, medina2019geometry, macedo2023shape}, MetaballVAE does not require prior knowledge of particle morphologies or shape-feature distributions as input, but only XRCT images of different sizes. Despite this, the satisfactory comparison results show that MetaballVAE is capable of accurately capturing overall shape features from parental XRCT particles and cloning new, morphologically similar avatars with similar feature distributions. Furthermore, this algorithm is effective even with small datasets of only a few dozen samples, which is often the case in engineering problems.

Besides, the above system is able to learn continuously. XRCT images can be kept added to the training dataset to improve the model performance. After short training (5 - 10mins), new morphological characteristics and shape-feature distributions can be learned by the original models. 





\subsection{Latent space arithematic}
A well-trained MetaballVAE can achieve integration of discrete training samples into continuous, regularized latent space, where the number of generated particles can be infinite. Such regularized latent space avoids overfitting and underfitting problems in generation, ensuring the rationality of generated morphologies\cite{shi2021randomly}. More importantly, it also realizes a certain control on the generated geometric feature\cite{zamorski2020adversarial}.  

\begin{figure}
    \centering
    \includegraphics[width=0.7\linewidth]{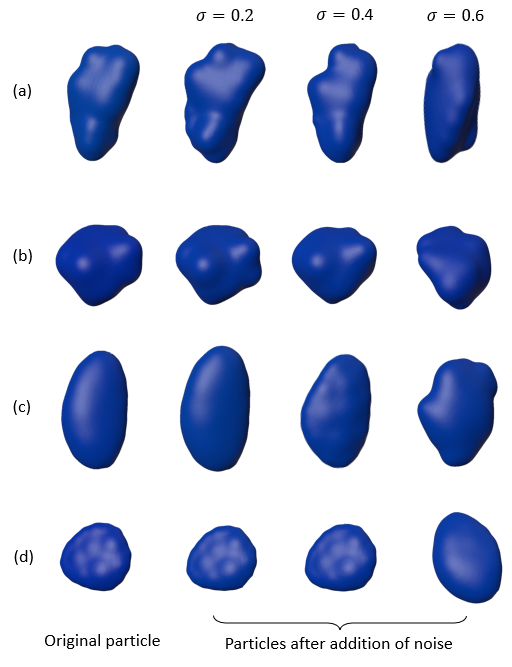}
    \caption{Comparison of generated particles before and after addition of Gaussian noise in the latent variables. (a) and (b) are Ottawa sand examples. (c) and (d) are cobblestone examples. The variance of Gaussian noise is noted as $\sigma$}
    \label{fig:adding_noise}
\end{figure}

Figure \ref{fig:adding_noise} illustrates that the addition of Gaussian random noises $\delta$ into the latent variable $\boldsymbol{z}$ can introduce morphological changes of different degrees into generated avatars. Here, $\delta$ is set to have the same dimension of $\boldsymbol{z}$, with zero mean value and different variance $\sigma$. Then, $\delta$s are added to a randomly selected latent variable $\boldsymbol{z}$ to produce modified ones $\boldsymbol{z}$ + $\delta$. Finally, these latent variables are fed into the generator as inputs. From the corresponding generated results, it can be observed that the addition of $\delta$s with small $\sigma$ can slightly adjust the particle morphologies. As the increase in $\sigma$, the degree of modification becomes larger, resulting in less similar avatars to the original one. This is from the property of regularized latent space. The addition of $\delta$s can create new latent variable adjacent to the original one, while the $\sigma$ of $\delta$s decides distances between them in the latent space. Since the latent space is regularized, such adjacent relationships control the shape similarity in generated avatars. This phenomenon is also observed in generating digital sand particles with VAE\cite{shi2021randomly}. It can be very useful when particles in certain morphologies are needed in simulations. We can first select the template avatars and then add $\delta$s of small magnitude into its latent variables. In this way, slightly modified avatars can be generated, avoiding repetitive particle morphologies in simulation. 

\begin{figure}
    \centering
    \includegraphics[width=0.7\linewidth]{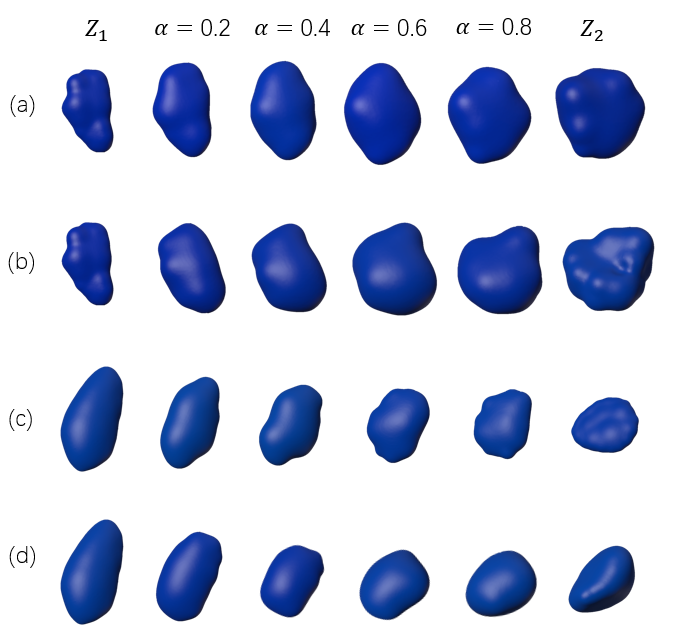}
    \caption{Comparison of generated particles of interpolated latent variables. (a) and (b) are cobblestone examples of the same $z1$. (c) and (d) are Ottawa sand examples of the same $z_1$. The interpolated coefficient is noted as $\alpha$. }
    \label{fig:interpolation}
\end{figure}

Figure \ref{fig:interpolation} indicates that interpolation between latent variables can produce smooth shape transitions in corresponding generated avatars. In these examples, two latent variable $z_1$ and $z_2$ are randomly selected to create interpolations $z_1$ + $\alpha(z_2 - z_1)$. Then, these latent variables are fed as input to the generator. It is clear that as the increase of $\alpha$, those interpolated avatars gradually transform from $z_1$ avatar to $z_2$ avatar. Note that such a change occurs simultaneously in multiple characteristics including shape, volume and surface area. This is also result of the regularized latent space. Those interpolated variables possess adjacent locations in the latent space with $z_1$ and $z_2$, resulting in avatars of similar shape. With change in the location of latent variables, the generated avatars show smooth modification in shape from $z_1$ to $z_2$ avatars. This phenomenon can be applied to obtain avatars of combined features. We can first select two template avatars of specific morphologies then do interpolation between them. 

\begin{figure}
    \centering
    \includegraphics[width=0.7\linewidth]{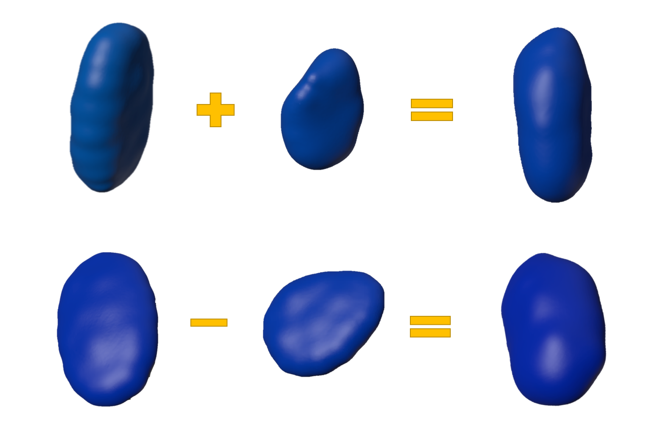}
    \caption{Manipulation of particle shape through addition and subtraction in latent variables. Top: adding angled features into a smooth cobblestone. Bottom: removing flattening features from a thin cobblestone. }
    \label{fig:arithematic}
\end{figure}

Figure \ref{fig:arithematic} shows that the avatar shape can be modified by applying addition or subtraction in the latent space. Here, arithmetic operations are implemented on latent variables, $z_1$ and $z_2$, corresponding to avatars of distinct shapes. Under such operation, specific shape features can be added or removed from the generated avatar of $z_3$. This also results from regularized latent space. The proper mapping between latent variables and morphological features provides a powerful method to modify the shape. Through proper arithmetic operation, avatars of specific features can be generated according to need. 

\section{Conclusions}

(1) We propose a geometric-based Metaball-Imaging (MI) algorithm, which can transform X-Ray Computed Tomography (XRCT) images of real particles into Metaball-based avatars. 

(2) The impact of the core hyperparameter, the number of control points, on particle characterization was carefully investigated. It was discovered that the characterization fidelity enhance with the increase of control-point number until reaching equilibrium.  For major morphological features of common particles, 40 control points are sufficient to do characterization. As for particles with more complex shapes, such as angular or concave features, only 120 control points are required to accurately represent these features.

(3) By comparing four types of commonly-encountered particles with distinct characteristics, the efficacy and precision of Metaball imaging (MI) in particle characterization are verified. The Metaball descriptor has been found to excel at depicting granular materials with round, smooth characteristics, such as cobblestones. The use of a lightweight Metaball avatar with only a handful of control points has been shown to be sufficient for creating high-fidelity representations of such particles. Overall, the findings of this evaluation suggest that MI is a valuable tool for accurately characterizing particles with distinct features.

(4) We also present a variational-autoencoder (VAE) based particle generation algorithm called MetaballVAE. It can generate style-similar particles in Metaball function form with XRCT images of parental particles. 

(5) The MetaballVAE was evaluated through a comparison of two groups of particles with different sizes. It was found that the parental and cloned particles exhibited good agreement in terms of their morphologies and shape-feature distributions. These results provide evidence that MetaballVAE is a reliable and practical tool for characterizing particles with varying sizes and morphologies.

(6) The regularized latent space of MetaballVAE allows for control over the generation process. Particles with specific morphologies can be generated through arithmetic operations on the latent space. This feature makes MetaballVAE a versatile and useful tool for generating particles with desired characteristics.


\section*{Acknowledgement}
We gratefully acknowledge the funds from National Natural Science Foundation of China (Project No.12172305), Natural Science Foundation of Zhejiang Province, China (LHZ21E090002), Key Research and Development Program of Zhejiang Province (Grant No.2021C02048) and Westlake University. We also thank Westlake High-performance Computing Center for computational sources and corresponding assistance.

\appendix
\section*{Appendix A. The theory of VAE}\label{VAE_deduction}
VAE is designed to learn the distribution of the input data $x$, which is controlled by some unobserved latent variable $\boldsymbol{z}$:
\begin{equation}
    P(x) = \int P(x, z) d z
\end{equation}

Since the above distribution is hard to learn directly for high dimensional problems, VAE approximates $P(x)$ with an assumed distribution $Q_\phi(x)$. The parameter $\phi$ is obtained through minimzing the KL divergence\cite{} between the real joint distribution $P(x, z)$ and the assumed one $Q_\phi(x, z)$:
\begin{equation}
    \arg \min _{\phi} KL(Q_\phi(x, z)||P(x, z))\label{VAE_o}
\end{equation}

The above divergence can be rewritten as:
\begin{equation}
\begin{aligned}
 KL(P(x, z) \| Q(x, z)) &= \mathbb{E}_{x \sim {P}(x)}[\ln {P}(x)] + \mathbb{E}_{x \sim {P}(x)}\left[\int P(z \mid x) \ln \frac{P(z \mid x)}{Q(x, z)} d z\right]
\end{aligned}
\end{equation}
where the first item is a constant. Then, a lower bound $\mathcal{L}$ of this divergence, which is called the evidence lower bound in variational inference, can be obtained:
\begin{equation}
\begin{aligned}
\mathcal{L} &= KL(Q_\phi(x, z)||P(x, z)) - constant \\
&=\mathbb{E}_{x \sim {P}(x)}\left[\int P(z \mid x) \ln \frac{P(z \mid x)}{Q(x, z)} d z\right]
\end{aligned}\label{ELBO1}
\end{equation}

This transforms the optimization objective(Eq. \ref{VAE_o}) into minimizing the above $\mathcal{L}$. 

To simplify the calculation of $\mathcal{L}$, we further apply the variational inference: 
\begin{equation}
\begin{aligned}
\mathcal{L} &= \underbrace{\mathbb{E}_{{x} \sim {P}(x)}\left[\mathbb{E}_{z \sim P(z \mid x)}[-\ln Q(x \mid z)]\right]}_{\text {Reconstruction Item}}
+ \underbrace{\mathbb{E}_{{x} \sim {P}(x)}{[KL(P(z \mid x) \| Q(z))]}}_{\text {Distribution Item}}
\end{aligned}\label{ELBO2}
\end{equation}

This joint expression consists of tow items: the reconstruction item and the distribution item. The first item evaluates the reconstructive performance of the input and the second item measures the distribution similarity between the generation and the input.

\section*{Appendix B. Derivation of the loss function}\label{metaballVAE_deduction}
The form of loss function for MetaballVAE is determined from the evidence lower bound of VAE(Eq. \ref{ELBO2}). In this equation, there are three undecided items: $P(z)$, $Q(x\mid z)$, and $P(z \mid x)$. We assume $\boldsymbol{z}$ to be multivariable normal distribution for regularization, which provides controls on the generated shape.  

In present study, $x$ refers to the shape representation, Metaball descriptor. Thus, Gaussian distribution is taken as the distribution of $Q(x \mid z)$. A neural network is adopted to represent its average $\mu(z)$ and variance $\sigma(z)$ with the reparameterizator, which gives:
\begin{equation}
Q(x \mid z)=\frac{1}{\prod_{k=1}^D \sqrt{2 \pi {\tilde\sigma}_{(k)}^2(z)}} \exp \left(-\frac{1}{2}\left\|\frac{x-{\tilde\mu}(z)}{{\tilde\sigma}(z)}\right\|^2\right)
\end{equation}
where $D$ is the dimension of latent variable $\boldsymbol{z}$. 

This allows $-\ln Q(x \mid z)$ to be transformed into:
\begin{equation}
    -\ln Q(x \mid z)=\frac{1}{2}\left\|\frac{x-{\tilde\mu}(z)}{{\tilde\sigma}(z)}\right\|^2+\frac{D}{2} \ln 2 \pi+\frac{1}{2} \sum_{k=1}^D \ln {\tilde\sigma}_{(k)}^2(z)\label{reconstruction1}
\end{equation}

In actual implementations, the variance $\sigma_(z)$ is taken as a constant. This simplifies Eq. \ref{reconstruction1} into:
\begin{equation}
    -\ln Q(x \mid z) \sim \frac{1}{2 {\sigma}^2}\|x-{\tilde\mu}(z)\|^2 \label{mse_loss}
\end{equation}

The above deductions enables approximation of the Reconstruction item with Mean Square Error form:
\begin{equation}
\mathbb{E}_{{x} \sim {P}(x)}\left[\mathbb{E}_{z \sim P(z \mid x)}[-\ln Q(x \mid z)]\right] \sim \frac{1}{d} \sum_{k=1}^d {\|x - \hat{x}\|}^2
\end{equation}
where the $d$ stands for the dimension of input $x$. 

We also assume $P(z \mid x)$ to be Gaussian. Similarly, a neural network is adopted to decide it:
\begin{equation}
    P(z \mid x)=\frac{1}{\prod_{k=1}^d \sqrt{2 \pi {\sigma}_{(k)}^2(x)}} \exp \left(-\frac{1}{2}\left\|\frac{z-{\mu}(x)}{{\sigma}(x)}\right\|^2\right)
\end{equation}
where ${\mu}(x)$ and ${\sigma}(x)$ are the corresponding average and variance. This allows the Distribution item to be expressed as:
\begin{equation}
K L(p(z \mid x) \| q(z))=\frac{1}{2} \sum_{k=1}^d\left(\mu_{(k)}^2(x)+\sigma_{(k)}^2(x)-\ln \sigma_{(k)}^2(x)-1\right)
\end{equation}

\bibliographystyle{elsarticle-num} 
\bibliography{mybib}

\begin{thebibliography}{10}
\expandafter\ifx\csname url\endcsname\relax
  \def\url#1{\texttt{#1}}\fi
\expandafter\ifx\csname urlprefix\endcsname\relax\def\urlprefix{URL }\fi
\expandafter\ifx\csname href\endcsname\relax
  \def\href#1#2{#2} \def\path#1{#1}\fi

\bibitem{meng2020three}
Q.~Meng, H.~Wang, M.~Cai, W.~Xu, X.~Zhuang, T.~Rabczuk, Three-dimensional
  mesoscale computational modeling of soil-rock mixtures with concave
  particles, Engineering Geology 277 (2020) 105802.

\bibitem{golombek2020geology}
M.~Golombek, N.~H. Warner, J.~A. Grant, E.~Hauber, V.~Ansan, C.~M. Weitz,
  N.~Williams, C.~Charalambous, S.~A. Wilson, A.~DeMott, et~al., Geology of the
  insight landing site on mars, Nature communications 11~(1) (2020) 1014.

\bibitem{zhang2021improved}
Y.~Zhang, J.~Shao, Z.~Liu, C.~Shi, An improved hydromechanical model for
  particle flow simulation of fractures in fluid-saturated rocks, International
  Journal of Rock Mechanics and Mining Sciences 147 (2021) 104870.

\bibitem{tolomeo2022modelling}
M.~Tolomeo, G.~R. McDowell, Modelling real particle shape in dem: a comparison
  of two methods with application to railway ballast, International Journal of
  Rock Mechanics and Mining Sciences 159 (2022) 105221.

\bibitem{altuhafi2016effect}
F.~N. Altuhafi, M.~R. Coop, V.~N. Georgiannou, Effect of particle shape on the
  mechanical behavior of natural sands, Journal of Geotechnical and
  Geoenvironmental Engineering 142~(12) (2016) 04016071.

\bibitem{santamarina2004soil}
J.~Santamarina, G.-C. Cho, Soil behaviour: The role of particle shape, in:
  Advances in geotechnical engineering: The Skempton conference: Proceedings of
  a three day conference on advances in geotechnical engineering, organised by
  the Institution of Civil Engineers and held at the Royal Geographical
  Society, London, UK, on 29--31 March 2004, Thomas Telford Publishing, 2004,
  pp. 604--617.

\bibitem{zhao2021grain}
X.~Zhao, D.~Elsworth, Y.~He, W.~Hu, T.~Wang, A grain texture model to
  investigate effects of grain shape and orientation on macro-mechanical
  behavior of crystalline rock, International Journal of Rock Mechanics and
  Mining Sciences 148 (2021) 104971.

\bibitem{lu2019re}
Z.~Lu, A.~Yao, A.~Su, X.~Ren, Q.~Liu, S.~Dong, Re-recognizing the impact of
  particle shape on physical and mechanical properties of sandy soils: a
  numerical study, Engineering Geology 253 (2019) 36--46.

\bibitem{chen2020effect}
R.-P. Chen, Q.-W. Liu, H.-N. Wu, H.-L. Wang, F.-Y. Meng, Effect of particle
  shape on the development of 2d soil arching, Computers and Geotechnics 125
  (2020) 103662.

\bibitem{shinohara2000effect}
K.~Shinohara, M.~Oida, B.~Golman, Effect of particle shape on angle of internal
  friction by triaxial compression test, Powder technology 107~(1-2) (2000)
  131--136.

\bibitem{xiao2019effect}
Y.~Xiao, Z.~Yuan, J.~Lin, J.~Ran, B.~Dai, J.~Chu, H.~Liu, Effect of particle
  shape of glass beads on the strength and deformation of cemented sands, Acta
  Geotechnica 14~(6) (2019) 2123--2131.

\bibitem{zhang2022size}
H.~Zhang, X.~Zhang, G.~Zhang, K.~Dong, X.~Deng, X.~Gao, Y.~Yang, Y.~Xiao,
  X.~Bai, K.~Liang, et~al., Size, morphology, and composition of lunar samples
  returned by chang’e-5 mission, Science China Physics, Mechanics \&
  Astronomy 65 (2022) 1--8.

\bibitem{zuo2019experimental}
L.~Zuo, S.~D. Lourenco, B.~A. Baudet, Experimental insight into the particle
  morphology changes associated with landslide movement, Landslides 16 (2019)
  787--798.

\bibitem{yin2020effect}
Z.-Y. Yin, P.~Wang, F.~Zhang, Effect of particle shape on the progressive
  failure of shield tunnel face in granular soils by coupled fdm-dem method,
  Tunnelling and Underground Space Technology 100 (2020) 103394.

\bibitem{zhou2020study}
B.~Zhou, D.~Wei, Q.~Ku, J.~Wang, A.~Zhang, Study on the effect of particle
  morphology on single particle breakage using a combined finite-discrete
  element method, Computers and Geotechnics 122 (2020) 103532.

\bibitem{cundall1979discrete}
P.~A. Cundall, O.~D. Strack, A discrete numerical model for granular
  assemblies, geotechnique 29~(1) (1979) 47--65.

\bibitem{iwashita1998rolling}
K.~Iwashita, M.~Oda, Rolling resistance at contacts in simulation of shear band
  development by dem, Journal of engineering mechanics 124~(3) (1998) 285--292.

\bibitem{wensrich2012rolling}
C.~Wensrich, A.~Katterfeld, Rolling friction as a technique for modelling
  particle shape in dem, Powder Technology 217 (2012) 409--417.

\bibitem{coetzee2016calibration}
C.~Coetzee, Calibration of the discrete element method and the effect of
  particle shape, Powder Technology 297 (2016) 50--70.

\bibitem{jiang2005novel}
M.~Jiang, H.-S. Yu, D.~Harris, A novel discrete model for granular material
  incorporating rolling resistance, Computers and Geotechnics 32~(5) (2005)
  340--357.

\bibitem{ferellec2010modelling}
J.~Ferellec, G.~McDowell, Modelling realistic shape and particle inertia in
  dem, G{\'e}otechnique 60~(3) (2010) 227--232.

\bibitem{hohner2012numerical}
D.~H{\"o}hner, S.~Wirtz, V.~Scherer, A numerical study on the influence of
  particle shape on hopper discharge within the polyhedral and multi-sphere
  discrete element method, Powder technology 226 (2012) 16--28.

\bibitem{regueiro2014micromorphic}
R.~Regueiro, B.~Zhang, F.~Shahabi, K.~Soga, Micromorphic continuum stress
  measures calculated from three-dimensional ellipsoidal discrete element
  simulations on granular media, IS-Cambridge 2014 (2014) 1--6.

\bibitem{galindo2013coupled}
S.~Galindo-Torres, A coupled discrete element lattice boltzmann method for the
  simulation of fluid--solid interaction with particles of general shapes,
  Computer Methods in Applied Mechanics and Engineering 265 (2013) 107--119.

\bibitem{grabowski2021comparative}
A.~Grabowski, M.~Nitka, J.~Tejchman, Comparative 3d dem simulations of
  sand--structure interfaces with similarly shaped clumps versus spheres with
  contact moments, Acta Geotechnica 16~(11) (2021) 3533--3554.

\bibitem{garcia2009clustered}
X.~Garcia, J.-P. Latham, J.-s. XIANG, J.~Harrison, A clustered overlapping
  sphere algorithm to represent real particles in discrete element modelling,
  Geotechnique 59~(9) (2009) 779--784.

\bibitem{thomas1995use}
M.~Thomas, R.~Wiltshire, A.~Williams, The use of fourier descriptors in the
  classification of particle shape, Sedimentology 42~(4) (1995) 635--645.

\bibitem{mollon2013generating}
G.~Mollon, J.~Zhao, Generating realistic 3d sand particles using fourier
  descriptors, Granular Matter 15~(1) (2013) 95--108.

\bibitem{zhou2018three}
B.~Zhou, J.~Wang, H.~Wang, Three-dimensional sphericity, roundness and fractal
  dimension of sand particles, G{\'e}otechnique 68~(1) (2018) 18--30.

\bibitem{zhou2015micromorphology}
B.~Zhou, J.~Wang, B.~Zhao, Micromorphology characterization and reconstruction
  of sand particles using micro x-ray tomography and spherical harmonics,
  Engineering geology 184 (2015) 126--137.

\bibitem{su20183d}
D.~Su, W.~Yan, 3d characterization of general-shape sand particles using
  microfocus x-ray computed tomography and spherical harmonic functions, and
  particle regeneration using multivariate random vector, Powder Technology 323
  (2018) 8--23.

\bibitem{vlahinic2017computed}
I.~Vlahini{\'c}, R.~Kawamoto, E.~And{\`o}, G.~Viggiani, J.~E. Andrade, From
  computed tomography to mechanics of granular materials via level set bridge,
  Acta Geotechnica 12~(1) (2017) 85--95.

\bibitem{kawamoto2016level}
R.~Kawamoto, E.~And{\`o}, G.~Viggiani, J.~E. Andrade, Level set discrete
  element method for three-dimensional computations with triaxial case study,
  Journal of the Mechanics and Physics of Solids 91 (2016) 1--13.

\bibitem{medina2019geometry}
D.~A. Medina, A.~X. Jerves, A geometry-based algorithm for cloning real grains
  2.0, Granular Matter 21~(1) (2019) 1--15.

\bibitem{zhao2019poly}
S.~Zhao, J.~Zhao, A poly-superellipsoid-based approach on particle morphology
  for dem modeling of granular media, International Journal for Numerical and
  Analytical Methods in Geomechanics 43~(13) (2019) 2147--2169.

\bibitem{zhao2020universality}
S.~Zhao, J.~Zhao, N.~Guo, Universality of internal structure characteristics in
  granular media under shear, Physical Review E 101~(1) (2020) 012906.

\bibitem{zhang2021metaball}
P.~Zhang, Y.~Dong, S.~Galindo-Torres, A.~Scheuermann, L.~Li, Metaball based
  discrete element method for general shaped particles with round features,
  Computational Mechanics 67~(4) (2021) 1243--1254.

\bibitem{zhang2022coupled}
P.~Zhang, L.~Qiu, S.~Galindo-Torres, Y.~Chen, A.~Scheuermann, L.~Li, Coupled
  metaball discrete element lattice boltzmann method for fluid-particle systems
  with non-spherical particle shapes: A sharp interface coupling scheme, arXiv
  preprint arXiv:2206.11634 (2022).

\bibitem{zhao2022metaball}
Y.~Zhao, P.~Zhang, L.~Lei, S.~Galindo-Torres, S.~Z. Li, Metaball-imaging
  discrete element lattice boltzmann method for fluid-particle system of
  complex morphologies with settling case study, arXiv preprint
  arXiv:2209.10411 (2022).

\bibitem{nie2020probabilistic}
J.-Y. Nie, D.-Q. Li, Z.-J. Cao, B.~Zhou, A.-J. Zhang, Probabilistic
  characterization and simulation of realistic particle shape based on sphere
  harmonic representation and nataf transformation, Powder Technology 360
  (2020) 209--220.

\bibitem{liu2011spherical}
X.~Liu, E.~Garboczi, M.~Grigoriu, Y.~Lu, S.~T. Erdo{\u{g}}an, Spherical
  harmonic-based random fields based on real particle 3d data: improved
  numerical algorithm and quantitative comparison to real particles, Powder
  Technology 207~(1-3) (2011) 78--86.

\bibitem{grigoriu2006spherical}
M.~Grigoriu, E.~Garboczi, C.~Kafali, Spherical harmonic-based random fields for
  aggregates used in concrete, Powder Technology 166~(3) (2006) 123--138.

\bibitem{wei2018generation}
D.~Wei, J.~Wang, J.~Nie, B.~Zhou, Generation of realistic sand particles with
  fractal nature using an improved spherical harmonic analysis, Computers and
  Geotechnics 104 (2018) 1--12.

\bibitem{xiong2021gene}
W.~Xiong, J.~Wang, Gene mutation of particle morphology through spherical
  harmonic-based principal component analysis, Powder Technology 386 (2021)
  176--192.

\bibitem{zhou2017generation}
B.~Zhou, J.~Wang, Generation of a realistic 3d sand assembly using x-ray
  micro-computed tomography and spherical harmonic-based principal component
  analysis, International Journal for Numerical and Analytical Methods in
  Geomechanics 41~(1) (2017) 93--109.

\bibitem{mollon2012fourier}
G.~Mollon, J.~Zhao, Fourier--voronoi-based generation of realistic samples for
  discrete modelling of granular materials, Granular matter 14~(5) (2012)
  621--638.

\bibitem{mollon20143d}
G.~Mollon, J.~Zhao, 3d generation of realistic granular samples based on random
  fields theory and fourier shape descriptors, Computer Methods in Applied
  Mechanics and Engineering 279 (2014) 46--65.

\bibitem{chen2022modified}
J.~Chen, R.~Li, P.-Q. Mo, G.~Zhou, S.~Cai, D.~Chen, A modified method for
  morphology quantification and generation of 2d granular particles, Granular
  Matter 24~(1) (2022) 1--18.

\bibitem{buarque2018granular}
R.~Buarque~de Macedo, J.~P. Marshall, J.~E. Andrade, Granular object
  morphological generation with genetic algorithms for discrete element
  simulations, Granular Matter 20~(4) (2018) 1--12.

\bibitem{shi2021randomly}
J.-j. Shi, W.~Zhang, W.~Wang, Y.-h. Sun, C.-y. Xu, H.-h. Zhu, Z.-x. Sun,
  Randomly generating three-dimensional realistic schistous sand particles
  using deep learning: Variational autoencoder implementation, Engineering
  Geology 291 (2021) 106235.

\bibitem{lai2022signed}
Z.~Lai, S.~Zhao, J.~Zhao, L.~Huang, Signed distance field framework for unified
  dem modeling of granular media with arbitrary particle shapes, Computational
  Mechanics (2022) 1--21.

\bibitem{jerves2017geometry}
A.~X. Jerves, R.~Y. Kawamoto, J.~E. Andrade, A geometry-based algorithm for
  cloning real grains, Granular Matter 19~(2) (2017) 1--10.

\bibitem{macedo2023shape}
R.~Macedo, S.~Monfared, K.~Karapiperis, J.~Andrade, What is shape?
  characterizing particle morphology with genetic algorithms and deep
  generative models, Granular Matter 25~(1) (2023) 1--12.

\bibitem{bourilkov2019machine}
D.~Bourilkov, Machine and deep learning applications in particle physics,
  International Journal of Modern Physics A 34~(35) (2019) 1930019.

\bibitem{kingma2013auto}
D.~P. Kingma, M.~Welling, Auto-encoding variational bayes, arXiv preprint
  arXiv:1312.6114 (2013).

\bibitem{higgins2016beta}
I.~Higgins, L.~Matthey, A.~Pal, C.~Burgess, X.~Glorot, M.~Botvinick,
  S.~Mohamed, A.~Lerchner, beta-vae: Learning basic visual concepts with a
  constrained variational framework (2016).

\bibitem{tan2018variational}
Q.~Tan, L.~Gao, Y.-K. Lai, S.~Xia, Variational autoencoders for deforming 3d
  mesh models, in: Proceedings of the IEEE conference on computer vision and
  pattern recognition, 2018, pp. 5841--5850.

\bibitem{blinn1982generalization}
J.~F. Blinn, A generalization of algebraic surface drawing. j-tog, 1 (3):
  235--256 (1982).

\bibitem{wang2022novel}
X.~Wang, Z.-Y. Yin, D.~Su, X.~Wu, J.~Zhao, A novel approach of random packing
  generation of complex-shaped 3d particles with controllable sizes and shapes,
  Acta Geotechnica 17~(2) (2022) 355--376.

\bibitem{plankers2001articulated}
R.~Plankers, P.~Fua, Articulated soft objects for video-based body modeling,
  in: Proceedings Eighth IEEE International Conference on Computer Vision. ICCV
  2001, Vol.~1, IEEE, 2001, pp. 394--401.

\bibitem{jin2000general}
X.~Jin, Y.~Li, Q.~Peng, General constrained deformations based on generalized
  metaballs, Computers \& Graphics 24~(2) (2000) 219--231.

\bibitem{bailey2004efficient}
D.~G. Bailey, An efficient euclidean distance transform, in: International
  workshop on combinatorial image analysis, Springer, 2004, pp. 394--408.

\bibitem{zhang2018improved}
Z.~Zhang, Improved adam optimizer for deep neural networks, in: 2018 IEEE/ACM
  26th International Symposium on Quality of Service (IWQoS), Ieee, 2018, pp.
  1--2.

\bibitem{wang2018optimization}
Y.~Wang, P.~Zhou, W.~Zhong, An optimization strategy based on hybrid algorithm
  of adam and sgd, in: MATEC Web of Conferences, Vol. 232, EDP Sciences, 2018,
  p. 03007.

\bibitem{sommer2011ilastik}
C.~Sommer, C.~Straehle, U.~Koethe, F.~A. Hamprecht, Ilastik: Interactive
  learning and segmentation toolkit, in: 2011 IEEE international symposium on
  biomedical imaging: From nano to macro, IEEE, 2011, pp. 230--233.

\bibitem{dietrich1982settling}
W.~E. Dietrich, Settling velocity of natural particles, Water resources
  research 18~(6) (1982) 1615--1626.

\bibitem{bouwman2004shape}
A.~M. Bouwman, J.~C. Bosma, P.~Vonk, J.~H.~A. Wesselingh, H.~W. Frijlink, Which
  shape factor (s) best describe granules?, Powder Technology 146~(1-2) (2004)
  66--72.

\bibitem{zhang2016lattice}
P.~Zhang, S.~Galindo-Torres, H.~Tang, G.~Jin, A.~Scheuermann, L.~Li, Lattice
  boltzmann simulations of settling behaviors of irregularly shaped particles,
  Physical Review E 93~(6) (2016) 062612.

\bibitem{mora2000sphericity}
C.~Mora, A.~Kwan, Sphericity, shape factor, and convexity measurement of coarse
  aggregate for concrete using digital image processing, Cement and concrete
  research 30~(3) (2000) 351--358.

\bibitem{para2021sketchgen}
W.~Para, S.~Bhat, P.~Guerrero, T.~Kelly, N.~Mitra, L.~J. Guibas, P.~Wonka,
  Sketchgen: Generating constrained cad sketches, Advances in Neural
  Information Processing Systems 34 (2021) 5077--5088.

\bibitem{zamorski2020adversarial}
M.~Zamorski, M.~Zi{\k{e}}ba, P.~Klukowski, R.~Nowak, K.~Kurach, W.~Stokowiec,
  T.~Trzci{\'n}ski, Adversarial autoencoders for compact representations of 3d
  point clouds, Computer Vision and Image Understanding 193 (2020) 102921.

\end{thebibliography}

\end{document}